\documentclass[12pt]{iopart}
\usepackage{graphicx}
\begin{document}
\title[Electron transfer in
multiwell nanostructures]{Resonant optical electron transfer in
one-dimensional multiwell structures}

\author{A V Tsukanov}

\address{Institute of Physics and Technology, Russian
Academy of Sciences, Nakhimovsky pr. 34, Moscow 117218, Russia}
\ead{tsukanov@ftian.ru}

\begin{abstract}
We consider coherent single-electron dynamics in the
one-dimensional nanostructure under resonant electromagnetic
pulse. The structure is composed of two deep quantum wells
positioned at the edges of structure and separated by a sequence
of shallow internal wells. We show that complete electron transfer
between the states localized in the edge wells through one of
excited delocalized states can take place at discrete set of times
provided that the pulse frequency matches one of resonant
transition frequencies. The transfer time varies from several tens
to several hundreds of picoseconds and depends on the structure
and pulse parameters. The results obtained in this paper can be
applied to the developments of the quantum networks used in
quantum communications and/or quantum information processing.
\end{abstract}
 \submitto{\JPC} \maketitle

\maketitle

\section{Introduction}

The implementation of reliable connection between remote parts of
quantum register remains one of the serious challenges on the way
of realization of scalable quantum computations. One possible
solution of this problem is to encode quantum information into the
spin or charge state of single probe electron that can next be
transferred from one qubit to another qubit with the help of an
auxiliary quantum circuit. In the theoretical schemes of
solid-state quantum information processing, such quantum networks
based on regular arrays of specific quantum nodes (semiconductor
quantum dots \cite{1}, quantum wells \cite{2}, quantum wires
\cite{3}, donor atoms \cite{4}), exploiting voltage-controlled
tunneling for coherent electron transport, were recently proposed.
The transfer protocols developed in the most of those works employ
the stimulated adiabatic passage techniques adopted from the
quantum optics. These methods were shown to be quite promising due
to their robustness against unwanted transitions from the
transport state. The electron transfer time depends on the
distance between the sender and target nodes as well as on the
voltage pulse design. In Ref. \cite{3} it was reported on the
possibility of complete electron transfer between two
electrostatically defined quantum dots connected through the 600
nm long quantum wire. The calculated transfer time $\tau\sim 50$
ps is short enough to allow coherent manipulations with the
nonlocal charge electronic state. However, since the voltage
driving schemes are very sensitive to imperfections in the time
delays and in the shapes of voltage pulses, the requirements on
the pulse engineering remain severe. This makes the realization of
such transfer protocols to be difficult experimental task. Other
strategies of selective electron transfer utilize the strong and
rapid electromagnetic modulation of the potential relief of
nanostructures \cite{5,6}. They rely upon the fact that the
tunneling rate between neighboring nodes in the structure can be
efficiently controlled by the field strength variation. At the
same time, the strong fields acting on the auxiliary network will
disturb the qubits in uncontrollable manner that may create
insurmountable obstacle on the route of scalable quantum
computer's building.

In this paper, we propose an alternative mechanism for the
selective electron transfer between two edge identical quantum
wells of the quasi-one-dimensional multiwell nanostructure, where
internal wells form the body of the transport circuit. The probe
electron, occupying initially the ground state of one of the edge
wells, can be brought into interaction with the nearest qubit.
Thus, the probe electron becomes entangled with this qubit, and
its spin or charge state is conditioned by the qubit state. Being
transported between two edge wells, the electron carries the
information about the state of a given qubit onto the qubit
attached to another edge well. Unlike conventional approaches
involving correlated voltage gate switches to operate on electron
charge state, we make use of $\Lambda$ - type resonant optical
excitation scheme proposed earlier to drive the single-qubit
evolution in various double-well quantum systems \cite{7} -
\cite{21}. At most two resonant terahertz pulses are required in
this case, and very accurate population transfer can be attained
in times of order of several tens of picoseconds and for the
pulses of much more moderate strengths than in the strategies
mentioned above. Here we study a way how to choose the pulse and
structure parameters in order to optimize the efficiency of
addressed probe electron transfer. We analyze the transfer
probabilities at different choices of the pulse strengths and
frequencies as well as for different numbers of quantum wells
forming the structure. Although the numerical results are obtained
for an effective one-dimensional structure, the algorithm of the
structure and pulse engineering developed here can be adopted for
realistic solid-state nanostructures.

The paper is organized as follows. In Sec. II the parameters of
individual quantum wells are selected in a way to form the
reliable single-electron transport channel in the nanostructure
composed of those wells. The energy spectrum and optical
properties of multiwell nanostructure are considered in Section
III. Section IV contains numerical data of the dynamical
simulations. We conclude our study by Section V.

\section{Single-well parameter choice}

We begin with the description of the model of nanostructure
employed in this work. The nanostructure is composed of $N$
quantum wells (QWs) placed in series and separated by finite
barriers. Two edge QWs with indices $i=1$ and $i=N$ are deeper
than the internal QWs. The energy spectrum of an electron confined
in the structure is presented by two ground states (each of them
being localized in corresponding edge QW) and $N$ excited
eigenstates delocalized over the structure. The latter result from
the hybridization of the states of individual QWs due to the
tunneling process. The localized states are supposed to be
isolated from each other, i. e., the electron tunneling between
them is negligibly small. If the electron occupies one of those
states, it can stay there for extremely long time. The excited
states of the nanostructure are approximately expressed by linear
combinations of the states of isolated QWs \cite{1}. Provided that
the nanostructure is symmetric, the wave functions of the
hybridized states have even or odd parity relative to the center
of nanostructure. The degree of hybridization depends on the
heights and the widths of the potential barriers separating
neighboring QWs as well as on the differences in the corresponding
eigenenergies of isolated QWs.

At the initial step, we choose the parameters of individual QWs.
Several circumstances should be taken into account here. First,
the ground states of the edge QWs are supposed to be strongly
localized whereas the excited states of those wells must be close
to the top of the potential barrier in order to form the reliable
transport channel (through their hybridization with the states of
neighboring internal QWs). Second, the matrix elements of optical
dipole transitions (ODTs) between the localized and excited states
of each edge QW have to be large enough to allow rapid optical
excitation/deexcitation of an electron. It is known that the value
of the ODT matrix element $\left| {d_{0,k} } \right|=\left|
{\langle 0 | ex |k \rangle } \right|$ ($e$ is the electron charge,
$x$ is the coordinate) between the ground state $|0 \rangle $ and
an arbitrary excited state $|k \rangle $ of a single QW goes down
quickly with the quantum number $k$ of the excited state. For a
square QW with infinite walls it depends on the excited state
number $k$ as $\left| {d_{0,k} } \right| \sim {{\left( {k + 1}
\right)} \mathord{\left/
 {\vphantom {{\left( {k + 1} \right)} {\left[ {k^2 \left( {k + 2} \right)^2 } \right]}}} \right.
 \kern-\nulldelimiterspace} {\left[ {k^2 \left( {k + 2} \right)^2 } \right]}}$,
where $k$ is odd because of the selection rule. It means that each
edge QW should be fabricated as a deep narrow well containing only
two bound electron states with $k=0$ and $k=1$. Further, since the
tunnel coupling between edge QWs is mediated by an array of
identical internal QWs, to implement electronic transfer over
longer distances, one has to utilize wider QWs and to increase the
QW number $N$. From the other hand, it is desirable that each
internal QW contain only one bound state. As a result, in order to
meet above conditions the internal QWs should be designed as wide
shallow wells.

Instead of commonly used square potential shape, we consider here
a less-studied power-exponential (PE) form for QW potentials
\cite{22}:
\begin{equation}
U\left( x \right) = U_0 \exp \left[ { - \left( {\frac{{x - x_0
}}{a_0}} \right)^{2p} } \right],
\end{equation}
where $U_0$ is the depth of potential well, $a_0$ is the radius of
potential, which determines the QW size, $x_0$ is the coordinate
of the center of potential, and parameter $p$ defines the
smoothness of QW boundaries. This kind of electron confinement is
found to be a good approximation for the electrostatically-gated
quantum dots in two- and three-dimensional cases \cite{22}. Wide
class of potentials can be modelled within this framework; in
particular, a square QW potential is obtained from Eq. (1) if one
sets $p>>1$. Note, that PE potentials are smooth functions of
coordinate that considerably simplifies the numerical treatment of
the eigenvalue problem.

The one-dimensional Schr\"odinger equation for a single electron
moving in the potential $U(x)$,
\begin{equation}
-\frac{{\hbar ^2 }}{{2m^* }}\frac{{\partial ^2 }}{{\partial x^2
}}\Psi \left( x \right) + U\left( x \right)\Psi \left( x \right) =
E\Psi \left( x \right),
\end{equation}
is solved with the help of the finite-difference second-order
scheme \cite{23}. This procedure is equivalent to the solution of
a system of uniform algebraic equations expressed by five-diagonal
sparse matrix and, consequently, to finding the eigenvalues and
eigenvectors of that matrix. The accuracy of the method is very
high, and the finite-difference schemes of higher orders are
generally excessive for treating the one-dimensional problems. The
results of calculations are given in effective atomic (or donor)
units - i. e., 1 a.u. = $Ry^* = m^* Ry/m_e \varepsilon ^2$ for the
energy and 1 a.u. = $a_B^* = m_e \varepsilon a_B /m^*$ for the
length, where $Ry$ is the Rydberg energy, $a_B$ is the Bohr
radius, $m_e$ is the free electron mass, $m^*$ is the effective
electron mass, and $\varepsilon$ is the dielectric constant. For
GaAs one has $Ry^*=0.006$ eV and $a_B^*$=10 nm.

Here we demonstrate the solutions of Eq. (2) for a single QW with
the PE potential of Eq. (1), where the potential radius $a_0$ and
the parameter $p$ are fixed and the potential depth $U_0$ is
varied. We used the computational interval of length $L$=10 with
the number of grid points $N_L$=10000 (at the boundaries of this
interval the wave functions are set to zero).  We take $a_0$ = 0.4
for the edge QW, $a_0$ = 1 for the internal QW, and $p$ = 5,
$x_0$=0 for both types of wells. The dependencies of low-lying
part of the energy spectrum for the edge and internal QWs on $U_0$
are presented in Figs. 1 (a) and 1 (b), respectively. We present
the results for the values of $U_0$ at which there are no more
than two bound electron eigenstates with energies
$\varepsilon_0<0$ and $\varepsilon_1<0$ in the QW. Other states,
whose energies $\varepsilon_k$ are positive (hereafter - unbound
or quasicontinuum states), belong to the spectrum of large square
infinite potential well of the length $L$, modified by the QW in
its central part. As the absolute value $|U_0|$ of the QW depth
decreases, the levels $\varepsilon_0$ and $\varepsilon_1$ shift to
the barrier top. They interact with unbound states of the same
symmetry giving rise to the anticrossing pattern. The presence of
the symmetric QW potential in the center of computation interval
amounts also to the energy splitting of states of the large well.
We observe apparent doublet structures for the states with
positive energies, which transform into the states of the large
square well at vanishing $U_0$.

Figures 2 (a) and 2 (b) illustrate the dependencies of absolute
values of ODT matrix elements $|d_{0,k}|$ on the potential depth
$U_0$. The values of the ODT matrix elements are given in atomic
units $|e|a_B^*$. We see that the plot of $|d_{0,1}|$ for the edge
QW [Fig. 2 (a)] has a local maximum at $U_0\approx -24$. At that
point, other matrix elements of ODT are still small, and all
excitations from the ground state of the edge QW to the
quasicontinuum states remain inefficient. It is thus reasonable to
take $U_0\approx$ -24 for the edge QW. As the QW becomes
shallower, one observes a sharp increase of the ODT matrix
elements, which approach the values of the ODT matrix elements
$d_{0,k}^{\left( L \right)}$ of the large square well. The latter
are much greater than the matrix elements of ODT between the bound
states of deep QW; the values for both types of matrix elements
relate each other as the corresponding effective potential
lengths, namely, ${|d_{0,k}^{\left( L \right)}/d_{0,k}|}\sim
{L/a_0}$, where $L/a_0\gg$ 1. At very large potential depths the
matrix element $d_{0,1}$ becomes close to the value
$d_{0,1}^{\left( 0 \right)} = {{16el_0 } \mathord{\left/
 {\vphantom {{16l_0 } {9\pi ^2 }}} \right.
 \kern-\nulldelimiterspace} {9\pi ^2 }}$ for an infinite square well, where
$l_0  = 2a_0 $ is the well width, and for $a_0$=0.4 one has
$d_{0,1}^{\left( 0 \right)}$ = 0.14. In contrast to the edge QW,
all of the ODT matrix elements of internal QWs should be selected
as small as possible in order to minimize unwanted excitations of
electron from bound state to the quasicontinuum states. As it is
seen in Fig. 2 (b), the plot of $|d_{0, 1}|$ demonstrates a
pronounced increase for $U_0\ge$ -2.5. Hence we choose -3 $\le U_0
\le$ -2.5 for the internal QW so that the first excited state is
outside the QW but the values of the ODT matrix elements are still
minor.

Making our choice of the value of $U_0$ for edge QWs [Fig. 1 (a)],
we fix the electronic eigenenergies $\varepsilon_0$ and
$\varepsilon_1$ of the edge QWs; then from the plot of the ground
state energy $\varepsilon_0$ of internal QW vs $U_0$ [Fig. 1 (b)]
we define the value of $U_0$ at which the energy $\varepsilon_0$
of internal QWs is approximately equal to the excited state energy
$\varepsilon_1$ of the edge QWs. The results of calculations
presented below are obtained for the following set of the
parameters: $a_0=$0.4, $U_0$ = - 23.83 for the edge QWs and
$a_0=$1, $U_0$ = - 2.5 for the internal QWs. From Figs. 1 (a) and
1 (b) one may see that the energy $\varepsilon_1$ of the edge QWs
slightly differs from the energy $\varepsilon_0$ of the internal
QWs. This is done intentionally in order to make our model of
nanostructure more realistic since in practice it is very
difficult to attain the exact adjustment of the QW levels.

\section{The nanostructure}

The $N$ QWs stacked in the way described above constitute the body
of a one-dimensional linear nanostructure. The energy spectrum of
the nanostructure depends on the parameters of individual QWs [the
QW potential depths $U_0(i)$ and the QW radii $a_0(i)$, $i$ = 1 -
$N$] and on the widths of the barriers separating the wells. For
simplicity, we suppose that all of the distances $b(i)$, $i$ = 1 -
($N-1$), between the centers of neighboring QWs (hereafter -
interwell distances) are the same and equal to $b$. In this case,
the potential of nanostructure is given by the expression
\begin{equation}
U\left( x \right) = \sum\limits_{i = 1}^N {U_0(i) \exp \left\{ { -
\left[ {\frac{{x - \left( {i - 1/2 - {N \mathord{\left/
 {\vphantom {N 2}} \right.
 \kern-\nulldelimiterspace} 2}} \right)b}}{{a_0(i)}}} \right]^{2p} }
 \right\}},
\end{equation}
where the origin of coordinate is placed into the center of
nanostructure. Now we seek the solution of Eq. (1) with the
potential having the form of Eq. (3) with the values of the edge
and internal QW parameters selected above and $p$=5. The
parameters for the edge (internal) QW are supplied with index 0
(1): $U_{0}(i)=U_{0(1)}$, $a_{0}(i)=a_{0(1)}$ if $i=1,N$
($i\ne1,N$).

In what follows, we shall be interested in eigenenergies of not
only the states bound in the QW structure, but also of states
lying higher than the potential barrier. The accurate calculation
of the energies of those quasicontinuum states requires a quite
large computational length $L$ and a great number $N_L$ of grid
points. We have computed the eigenenergies $\varepsilon _k$ in the
range $ \varepsilon _0 \le \varepsilon _k  \le \left| {\varepsilon
_0 } \right|$ for different QW numbers $N$ (up to $N$ = 20) at
several sets of $L$ and $N_L$. It was found that for all those
nanostructures the choice $L$ = 500 and $N_L$ = 40000 enables one
to satisfactory simulate the quasicontinuum states in the pointed
energy interval.

In this Section, we examine in detail the nanostructure composed
of six QWs, see Fig. 3 (a). The plots of the excited state
energies $\varepsilon _k$ ($k$=2 - 7) on the interwell distance
$b$ (actually, on the barrier width) are presented in Fig. 3 (b).
As expected, the energies $\varepsilon _k$ depend on $b$
exponentially. At the same time, they deviate from the
tight-binding (TB) predictions \cite{1} and bring about an
asymmetrical distribution of $\varepsilon _k$ relative to the
center of the energy subband. This is the peculiarity of a real
system, where the energy of a hybridized state is to be a function
of the closeness of this state to the barrier top. In fact, the
overall correspondence between the spectrum of our structure and
the spectrum of structure modelled within the TB approximation
\cite{1} is mainly determined by the effective tunneling rates
between neighboring QWs. Usually, it is expected that the weaker
the tunneling the better this correspondence. Therefore, for
deeper wells and wider barriers the results of TB model  will give
more accurate description of the nanostructure. Besides, the small
detuning in the energies of individual QWs affects, to some
extent, the mechanism of the state hybridization. For a quite
large $b$ ($b\ge$4.5), it results in a dissociation of the
six-fold energy subband into the two degenerate energy levels
($\varepsilon_2$ and $\varepsilon_3$), presented by the even and
odd superpositions of excited states of isolated edge QWs, and the
four-fold subband of weakly-hybridized states with
$\varepsilon_{k\ne 2,3}$ pertaining to the internal part of the
nanostructure. As follows from our calculations, the variation of
the distance $b_0$ between the centers of the edge QWs and the
neighboring internal QWs at some fixed distance $b_1$ between the
centers of two neighboring internal QWs does not change noticeably
the properties of the excited subband. The wave functions
$\Psi_k(x)$ of the bound states $\left| k \right\rangle $ ($k$=0 -
7) are shown in Fig. 4. Note, that the wave functions $\Psi_0(x)$
and $\Psi_1(x)$ forming the ground state subspace of the Hilbert
space of multiwell nanostructure are taken as the functions each
localized in corresponding edge QW. This representation given by
the isolated edge QW orbitals  is equivalent to the representation
given by their symmetric and antisymmetric combinations provided
that the corresponding eigenenergies $\varepsilon_0$ and
$\varepsilon_1$ of the Hamiltonian are almost degenerate.

Other important quantities, the ODT matrix elements, are plotted
in Figs. 5 (a) and 5 (b). Due to the inversion symmetry of our
structure, we may calculate the values of the matrix elements for
only one excitation arm of the $\Lambda$ scheme, e. g., involving
the ground state $\left| 0 \right\rangle $, localized in the left
QW ($i$=1), and the excited states $\left| k \right\rangle $
($k$=2 - 7). The ODT matrix elements for another excitation arm
with the ground state $\left| 1 \right\rangle $ of the right QW
($i$=$N$), are related to them by the formula $d_{1,k}=(-1)^k
d_{0,k}$. In the range of the interwell distances $b$ around
$b=$3, the values of matrix elements $|d_{0,k}|$ for pairs of the
states (doublets) $\{\left| 2 \right\rangle $, $\left| 7
\right\rangle \}$, $\{\left| 3 \right\rangle $, $\left| 6
\right\rangle \}$, and $\{\left| 4 \right\rangle $, $\left| 5
\right\rangle \}$ remain close to the corresponding values
calculated in the TB model (denoted by horizontal dotted lines).
As in the case of the energies, both data types do not coincide.
This discrepancy can be explained in the following way. The TB
wave functions of hybridized states are constructed from the
isolated QW wave functions multiplied by the weight coefficients
that depend on both the state number $k$ and the QW index $i$
\cite{1}. Consequently, the ODT matrix element between the ground
state $\left| 0 \right\rangle $  and the hybridized state $\left|
k \right\rangle $ is simply the single-well matrix element
$d_{0,1}$ [Fig. 2 (a)] multiplied by the weight coefficient for
the state $\left| k \right\rangle $ in the edge QW ($i$=1, $N$).
For two quantum states of a given doublet, those coefficients are
the same in one of the edge QWs and differ by the sign in another
one. Hence, the moduli of the ODT matrix elements $|d_{0 (1),k}|$
for doublet states turn out to be equal to each other and do not
depend on $b$ under the TB approximation. However, we observe in
Fig. 4 that the wave functions of doublet states have close but
still different amplitudes in the edge QW regions. As a result,
the matrix elements of ODT are different for the two doublet
states. It means also that the wave functions of
highly-delocalized states of the realistic nanostructure cannot be
satisfactory approximated by the linear combinations of the wave
functions of isolated QWs.

If the QWs are substantially remote from each other ($b\ge$ 4.5),
the curves for $|d_{0,2}|$ and $|d_{0,3}|$ demonstrate
asymptotical convergence to a common non-zero value whereas the
others decay slowly to zero. Such a behavior of the ODT matrix
elements indicates on the collapse of tunneling between the edge
QWs and the internal part of nanostructure. Since the states
$\left| 2 \right\rangle $ and $\left| 3 \right\rangle $ evolve
with $b$ into even and odd superpositions of excited states of the
isolated edge QWs, the linear combination ${{\left( {\left|
{d_{0,2} } \right| + \left| {d_{0,3} } \right|} \right)}
\mathord{\left/
 {\vphantom {{\left( {\left| {d_{0,2} } \right| + \left| {d_{0,3} } \right|} \right)} {\sqrt 2 }}} \right.
 \kern-\nulldelimiterspace} {\sqrt 2 }}$ tends to the value of $|d_{0,1}|$ for isolated edge QW.
Note, that all of  $|d_{0 ,k}|$, involving the quasicontinuum
states $\left| k \right\rangle $ ($k>$7), are very small in
comparison with the values of matrix elements for ODTs between the
bound states. [Several matrix elements for the transitions from
the ground state into the quasicontinuum states are shown as thin
black curves at the bottom of Fig. 5 (a)]. The explanation of this
fact is rather simple. In general, the one-dimensional wave
function of a localized electron may be presented in a form $\Psi
_k \left( x \right) = {{\tilde \Psi _k \left( x \right)}
\mathord{\left/
 {\vphantom {{\tilde \Psi _k \left( x \right)} {\sqrt {l_k } }}} \right.
 \kern-\nulldelimiterspace} {\sqrt {l_k } }}$, where $k$ is the quantum number of the electron state,
$l_k$ is the characteristic length of electron spreading in this
state, and $\tilde \Psi _k \left( x \right)$ is the dimensionless
function which amplitude is of order of unity. The ODT matrix
element between the states $\left| 0 \right\rangle $ and $\left| k
\right\rangle $ can thus be written as $d_{0,k}  = C_{0,k} e
{{l_0^2 } \mathord{\left/
 {\vphantom {{l_0^2 } {\sqrt {l_0 \,l_k } }}} \right.
 \kern-\nulldelimiterspace} {\sqrt {l_0 \,l_k } }}$, where
$|C_{0,k}|\le$1. Substituting here $l_{0}=l_{k}=a_0$, we see that
corresponding matrix element of the ODT, involving two bound
states $\left| 0 \right\rangle $ and $\left| k \right\rangle $, is
of the order of $ea_0$. If, however, one sets $l_{0}=a_0$ and
$l_{k}=L$, the estimation $\left| {d_{0,k} } \right| \sim ea_0
\sqrt {{{a_0 } \mathord{\left/
 {\vphantom {{a_0 } L}} \right.
 \kern-\nulldelimiterspace} L}}  \ll ea_0 $ is obtained.
Both results are consistent with the numerical data presented in
Fig. 5 (a).

Considering the next type of the matrix elements, we deal with the
transitions between the states that constitute one subband. In
Fig. 5 (b), we show the data illustrating the dependencies of such
intrasubband ODT matrix elements on interwell distance $b$ for the
subbands $\{\left| 0 \right\rangle $, $\left| 1 \right\rangle\} $
and $\{\left| 2 \right\rangle $ - $\left| 7 \right\rangle \}$. In
contrast to the intersubband ODT matrix elements $d_{0 ,k}$ ($k$ =
2 - 7) that vary smoothly with $b$, the intrasubband matrix
elements are approximately linear functions of the interwell
distance. We observe also a strong increase in the values of the
intrasubband ODT matrix elements in comparison with the
intersubband ones. Besides, the intrasubband ODT matrix elements
between neighboring states are significantly larger than those
between non-neighboring states, pictured as thin black curves in
Fig. 5 (b). According to the TB model, all intrasubband matrix
elements are expressed by linear combinations of the diagonal
single-well matrix elements. The latter describe the shifts of the
energy levels of isolated QWs in an electric field. Their
difference has a clear analogy in the classical physics: it is
just the energy acquired by an electron transported between the QW
centers in the electric field of unit strength. In the case of the
diagonal matrix element for the localized state $\left|
0\right\rangle$, the "classical" formula $|d_{0,0}| = {{b\left( {N
- 1} \right)} \mathord{\left/
 {\vphantom {{b\left( {N - 1} \right)} 2}} \right.
 \kern-\nulldelimiterspace} 2}$ perfectly reproduces  the plot for $|d_{0 ,0}|$ vs $b$ found
in the essentially quantum treatment of the problem. With that, as
$b$ increases, the plots of matrix elements for excited subband
substantially deviate from the linear form. For example, the
matrix element $d_{2, 3}$ initially separates from the others and
then approaches $d_{0 ,0}$ at $b\ge$ 4.5. Again, it is because of
the reorganization in the structure of hybridized states due to
the tunneling collapse. Thus, the intrasubband ODT matrix elements
not only account of the effects associated with the charge
displacement in the external field, but also reflect the
peculiarities of electronic density distribution in given quantum
states.

The eigenenergies $\varepsilon_n$ and matrix elements $d_{m,n}$ of
ODT calculated above will be used in the next Section in numerical
study of the dynamical problem.

\section{Dynamics}

We have mentioned in Introduction that there exist several ways of
how to transport an electron from one QW to another in a
semiconductor nanostructure. The aim of the present paper is to
clarify some new aspects of resonant optical excitations in the
effective $\Lambda$ system formed in the spatially extended
multiwell structure by a pair of ground states and a set of
delocalized states. In the work \cite{7}, the symmetric
three-level scheme was employed to describe the resonant electron
dynamics in the double-dot structure. It was shown that the square
harmonic laser pulse of strength $E\left( t \right)=E_0\cos\left(
{\omega t} \right)$ and duration $T$ can produce complete
population transfer between two spatially remote ground states
$\left| 0\right\rangle$ and $\left| 1\right\rangle$ of the
nanostructure if the pulse frequency $\omega$ is tuned exactly to
the resonance with the transitions $\left\{ {\left| 0
\right\rangle ,\left| 1 \right\rangle } \right\}
\mathbin{\lower.3ex\hbox{$\buildrel\textstyle\rightarrow\over
{\smash{\leftarrow}\vphantom{_{\vbox to.5ex{\vss}}}}$}} \left|
r\right\rangle $ connecting the ground state subspace and some
excited (transport) state $\left| r\right\rangle$ (other excited
states are ignored). The accuracy of such a three-level
approximation is very high provided that the following conditions
imposed on the structure and pulse parameters are satisfied. (i)
The transition frequency $\omega$ is much larger than both the
interlevel spacings
$|\Delta_{r,r\pm1}|=|\varepsilon_r-\varepsilon_{r\pm1}|$ and the
coupling coefficient $\lambda_{0,r}=E_0d_{0,r}/2$. It allows one
to apply the so-called rotating wave approximation (RWA) under
which an analytic solution of the dynamical problem is easily
found. (ii) The coupling coefficient $\lambda_{0,r}$ is rather
small as compared with $|\Delta_{r,r\pm1}|$. It guarantees the
selectivity of resonant excitations, i. e., only the transport
state $\left| r \right\rangle $ is populated during the pulse
action. (iii) The quantum dots (wells) are close enough to each
other to neglect the diagonal coupling coefficients. It means that
rapidly varying energy shifts do not modify appreciably the
transition frequency $\omega_{0(1),r}$ and thus do not violate the
resonant conditions. The correction to the bare three-level scheme
\cite{7}, accounting for a non-resonant excitation of the
delocalized state nearest to the transport state, as well as the
first-order correction to the RWA, were derived in Ref. \cite{10}.

Here we investigate the problem of the resonant field-structure
interaction via fully-numerical treatment. We solve the
one-electron non-stationary Schr\"odinger equation
\begin{equation}
i\hbar \frac{\partial }{{\partial t}}\left| {\Psi \left( t
\right)} \right\rangle  = H\left( t \right)\left| {\Psi \left( t
\right)} \right\rangle
\end{equation}
with the Hamiltonian
\begin{equation}
H\left( t \right) = H_0  - eE_0x \cos \left( {\omega t} \right),
\end{equation}
where $H_0$ is the Hamiltonian of unperturbed nanostructure with
the potential relief given by Eq. (3).

The state vector  of the system may be presented in the form
\begin{equation}
\left| {\Psi \left( t \right)} \right\rangle  = \sum\limits_{n
}^{} {c_n \left( t \right)e^{ - i{{\varepsilon _n t}
\mathord{\left/
 {\vphantom {{\varepsilon _n t} \hbar }} \right.
 \kern-\nulldelimiterspace} \hbar }} \left| n \right\rangle },
\end{equation}
where index $n$ runs over the states belonging to the low-energy
part of the nanostructure spectrum. Inserting Eqs. (5) and (6)
into Eq. (4) and then multiplying Eq. (4) from the left side by
$\langle m |$ ($m$ = 0, 1, 2,..), we arrive at the set of linear
differential equations for the probability amplitudes $c_m \left(
t \right)$. It is convenient to rewrite this set in the following
way:
\begin{equation}
i\frac{{\partial c_m }}{{\partial t}} = 2\sum\limits_{n }^{} {c_n
\lambda_{m,\,n} \cos \left( {\omega t} \right)e^{ - i\omega
_{m,\,n} t} } \,\,\,(m=0,\,1,\,2,\,..),
\end{equation}
where $\omega _{m,\,n}=\Delta
_{m,\,n}/\hbar=(\varepsilon_n-\varepsilon_m)/\hbar$ is the
transition frequency between the states $\left| m \right\rangle$
and $\left| n \right\rangle$, $\lambda_{m,\,n}=\varepsilon
_{field} {d_{m,\,n}}/2$ is the corresponding coupling coefficient,
$\varepsilon _{field} = {{ea_B^* E_0 } \mathord{\left/
 {\vphantom {{ea_B^* E_0 } {Ry^* }}} \right.
\kern-\nulldelimiterspace} {Ry^* }}$ is the field energy, and the
time $t$ is taken in units of $\hbar/Ry^*$ (for GaAs 1 a.u.=0.11
ps).

For definiteness, assume that at $t$=0 the electron is localized
in the state $\left| 0 \right\rangle $ in the left edge QW - i.
e., $\left| {\Psi \left( 0 \right)} \right\rangle =\left| 0
\right\rangle $, and hence the initial conditions for Eq. (7) read
$c_m \left( 0 \right)=\delta_{m,0}$. Since our attention is
concentrated on the possibility of complete population transfer
between the ground states $\left| 0 \right\rangle $ and $\left| 1
\right\rangle $ of the multiwell nanostructure, here we restrict
ourselves by analysis of the probability $p_1(T)=|c_1(T)|^2$ to
find the electron in the target state $\left| 1 \right\rangle $
after the pulse is switched off. The numerical solutions of Eq.
(7) have been obtained with the help of the non-stiff ode113
Matlab solver with relative accuracy $\delta_{tol}=10^{-6}$. In
all simulations we take the pulse frequency $\omega$ to be equal
to one of the nanostructure transition frequencies
$\omega_{0(1),\,r}$ (2 $\le r\le$ 7 for six-well structure). In
this case, the single-electron dynamics is given by the resonant
three-level Rabi oscillation picture. Typical plots of
probabilities $p_0(T)$ and $p_1(T)$ of finding an electron in the
states $\left| 0 \right\rangle $ and $\left| 1 \right\rangle $,
together with the total probability $p_{tr} \left( T \right) =
\sum\limits_{k \ne 0,1} {p_k \left( T \right)} $
[$p_k(T)=|c_k(T)|^2$] of electron to be out of ground-state
subspace, versus the pulse duration $T$, are presented in Fig. 6
for $r$ = 5, $N$ = 6.

The number of states, included into the simulations, was varied
from 30 to 40. Apart from the bound states shown in Fig. 3 (a), we
also took into account two groups of quasicontinuum states. The
first group contains the states with energies
$0<\varepsilon_k\ll|\varepsilon_0|$ lying in nearest neighborhood
of the barrier top. It is expected that those states may affect,
to some extent, the resonant dynamics in the bound state subspace
through non-resonant excitations. The second group incorporates
the states from the narrow energy interval located around the
energy $\varepsilon_r+\omega$. Since those states are close to the
two-photon resonance with the driving pulse, they could
participate the electron evolution despite of small values of the
corresponding ODT matrix elements. However, as one observes in
Fig. 6, the resonant dynamics of an electron in the multilevel
system is an essentially three-level dynamics. It is very close to
that considered in Ref. \cite{7} and is described by the formulae
\begin{equation}
p_0  = \cos ^4 \left( {\Omega _R t} \right),\,\,\,p_1  = \sin ^4
\left( {\Omega _R t} \right),\,\,\,p_r  = \frac{1}{2}\sin ^2
\left( {2\Omega _R t} \right),
\end{equation}
where $\Omega _R  = {{\left| {\lambda _{0,\,r} } \right|}
\mathord{\left/
 {\vphantom {{\left| {\lambda _{0,\,r} } \right|} {\sqrt 2 }}} \right.
 \kern-\nulldelimiterspace} {\sqrt 2 }}$ is the Rabi frequency.
It means that only the states $\left| 0 \right\rangle $, $\left| 1
\right\rangle $, and $\left| r \right\rangle $  play a decisive
role in the electron dynamics, whereas the presence of other
states is manifested in the weak modulations of ideal three-level
plots. The participation of ladder resonant excitations involving
quasicontinuum states as well as the processes beyond the RWA may
be safely ignored due to the smallness of the parameters ${{\left|
{\lambda _{r,\,k} } \right|} \mathord{\left/ {\vphantom {{\left|
{\lambda _{r,\,k} } \right|} {\left| {\lambda _{0,\,r} }
\right|}}} \right. \kern-\nulldelimiterspace} {\left| {\lambda
_{0,\,r} } \right|}}$ ($k>$ 7) and ${{\left| {\lambda _{0,\,r} }
\right|} \mathord{\left/ {\vphantom {{\left| {\lambda _{0,\,r} }
\right|} \omega }} \right. \kern-\nulldelimiterspace} \omega }$.
The transfer time $T_0$ is thus defined from the condition
$\Omega_R T_n=\pi/2+\pi n$ ($n$=0,\,1,\,2,..) as the shortest
pulse duration $T_n$ at which the maximal population of target
state $\left| 1 \right\rangle $ is achieved - i. e.,
$T_0=\pi/2\Omega_R$. Note, that the definition of Rabi frequency
$\Omega_R$ is taken according to Refs. \cite{7} and \cite{10}; it
differs by $1/\sqrt{2}$ from that used in Ref. \cite{11} for the
description of resonant hydrogen molecular ion dynamics [cf.
\cite{11}, Eq. (40)], since in that work the ground states of a
molecular ion, as well as its excited states, were presented by
even and odd linear combinations of the isolated atom states.

The most important question one has to answer is what a state from
the excited subband should be used as the transport state $\left|
r \right\rangle $ in order to optimize the transfer probability
$\max(p_1)$ and to reduce the transfer time $T_0$. According to
the selectivity and high speed requirements, the values of $|d_{0,
r}|$ and $|\Delta_{r, r\pm1}|$ should be taken as large as
possible. As it is clearly seen from both numerical data and
analytical TB estimations, the states from the center of the
excited energy subband (in the six-well structure the states
$\left| 4 \right\rangle $ and $\left| 5\right\rangle $) are
characterized by the largest interlevel spacings and, at the same
time, by the highest ODT matrix elements $|d_{0, r}|$ than the
states at the subband edges.   We have performed the calculations
of maximal transfer probabilities and transfer times at several
fixed field energies $\varepsilon_{field}$ (actually, the field
strength) and for the different field frequencies
$\omega=\omega_{0, k}$ ($k$=2 - 7). The results of calculations
confirm our expectations, demonstrating that
$T_{0,k=4}<T_{0,k=5}<T_{0,k=3}<T_{0,k=6}$ in the whole range of
$\varepsilon_{field}$. The pulse durations $T_{0,k=2}$ and
$T_{0,k=7}$ corresponding to the electron transfer via edge
excited subband states are significantly longer than others. Thus
we come to the conclusion that the state closest to the center of
excited subband is the best candidate for the transport state. If
the QW number $N$ is even, it will be one of states $\{\left|
N/2+1 \right\rangle $, $\left| N/2+2 \right\rangle \}$ of the
central doublet, whereas for odd $N$ this is the unpaired state
$\left| 1+(N+1)/2 \right\rangle $.

The plots of the maximum values of $p_1$, see Fig. 7, demonstrate
quasiperiodic behavior of the transfer probabilities on the field
energy for each state of the pair $\{\left| 4\right\rangle $,
$\left| 5 \right\rangle \}$ for $N$=6. We observe that the state
$\left| 4 \right\rangle $ displays better transport
characteristics in comparison with the state $\left| 5
\right\rangle $. The maximum of the transfer probability $\max
(p_1)\approx$ 0.999 for $\omega=\omega_{0,\, 4}$ is attained at
quite large field energy $\varepsilon_{field}$ = 0.083, thus
allowing for rapid electron transfer (about 45 ps for the GaAs
parameters). The transfer times $T_0$ for both
$\omega=\omega_{0,\, 4}$  and $\omega=\omega_{0,\,5}$ cases are
presented in Fig. 8. It is worth mentioning that the simple
three-level expression for $T_0$ (shown by the dashed curves) fits
the numerical graphs very well.

In order to explain the dependencies of $\max(p_1)$ on the field
energy, we make a natural assumption that the oscillations in
corresponding plots are caused by the presence of other excited
states close to the transport state $\left| r\right\rangle $. In
the work \cite{10}, we calculated the correction to ideal resonant
three-level dynamics [Eq. (8)] taking into account the
non-resonant occupation of the neighboring state $\left|
k\right\rangle $. The corrected four-level formula for the
population $p_1$ of target state $\left| 1\right\rangle $ at
$T=T_0$ reads in current notations
\begin{equation}
p_1 \left( {T_0 } \right) \approx 1 - f_{k } ,\,\,\,f_{k }  =
\left( {{{2\lambda _{0,\,k} } \mathord{\left/
 {\vphantom {{2\lambda _{0,\,k} } {\Delta _{k,\,r} }}} \right.
 \kern-\nulldelimiterspace} {\Delta _{k,r} }}} \right)^2 \sin ^2 \left( {{{\pi \Delta _{k,\,r} } \mathord{\left/
 {\vphantom {{\pi \Delta _{k,\,r} } {2\sqrt 2 \lambda _{0,\,r} }}} \right.
 \kern-\nulldelimiterspace} {2\sqrt 2 \lambda _{0,\,r} }}} \right).
\end{equation}
Note that this formula was derived via accurate solution of the
four-level dynamical problem and contains more profound physics
than those obtained by the simple adiabatic exclusion of the
non-resonant levels.

The form of the set of Eqs. (7) assures that the non-resonant
excitations of the delocalized nanostructure states  $\left|
k\right\rangle $ ($k{\ne}r$) occur independently up to the terms
of higher orders in the parameters $\lambda _{0,\,k}/\Delta
_{k,r}$, and the direct generalization of Eq. (9) to the
multilevel $\Lambda$ system results in the formula
\begin{equation}
\max \left( {p_1 } \right) = p_1 \left( {T_0 } \right) \approx 1 -
\sum\limits_{k \ne r} {f_{k} }
\end{equation}
(hereafter, the sum is over the bound states only). The
dependencies of $\max \left( {p_1 } \right) $, calculated through
the Eq. (10), are visualized in Fig. 7 as the dashed curves. We
see that for $\omega=\omega_{0, 4}$ the formula (10) is in a good
agreement with the numerical data, while for $\omega=\omega_{0,
5}$ it gives overestimated values for local maxima of the transfer
probability. What is the reason for such
 deviations? To our opinion, the main error originates from the
neglecting in $f_k=|c_k|^2$ [Eqs. (9) and (10)] the interference
terms describing the correlation effects between the state $\left|
k\right\rangle $ and other excited states. Despite of the small
values of these terms in comparison with non-correlated terms
entering the formula (10), the correlated excitations may amount
to some reduction of transfer probability. Besides, in the
derivation of Eq. (9), the diagonal (intrasubband) coupling
coefficients were ignored. These intrasubband terms of Eq. (7)
oscillate at the field frequency $\omega$ relative to the RWA
frame and are usually discarded. However, since their values grow
linearly with the interwell distance, they will affect the
electron dynamics in extended structures, as $\lambda_{m,\,n}$
become close to $\omega$.

We believe that the difference in the influence of one excited
state on another excited state is explained by the difference in
their ODT matrix elements $|d_{0, k}|$ . For example, the value of
$|d_{0, 4}|$ is larger than the value of $|d_{0, 5}|$; as a
result,  the population of the state $\left| 5 \right\rangle $ is
subjected to stronger modulations caused by the state $\left| 4
\right\rangle $, while the influence of the state $\left| 5
\right\rangle $ on the the population of the state $\left| 4
\right\rangle $ is less pronounced. It means that the state
$\left| 4 \right\rangle $ possesses better transport properties
than the state $\left| 5 \right\rangle $ and thus it should be
selected as the transport state in the six-well nanostructure. The
origin of oscillations of the transfer time $T_0$ may be
interpreted in a similar way. It should be noted that the
first-order oscillating correction to $T_0$ does not improve the
accuracy of Eq. (10); this observation agrees with the arguments
given above that several processes were neglected in derivation of
Eq. (10). Nevertheless, the analytical curves yield rather
accurate values of the local maxima positions in both cases.
Therefore, Eq. (10) may be used at initial step for evaluation of
the laser field strengths for which the probability of electron
transfer is maximal. On the other hand, this formula enables us to
study the dependence of $\max(p_1)$ on the energy splittings $
\Delta _{k,\,r} $ that, in their turn, are the functions of the
interwell distance $b$ and the QW number $N$.

The important feature in the behavior of $\max(p_1)$, expressed by
Eq. (10) and visualized in Fig. 9 for $N$ = 6, is its band-like
structure, where high transfer probability regions alternate with
low transfer probability regions. We observe that $\max(p_1)$
exhibits oscillations as $b$ varies. Numerical data, plotted in
Fig. 7 (a) for $b$ = 3, lie at the vertical dashed line in Fig. 9.
We have found that the value of $\max(p_1)$=0.999  at
$\varepsilon_{field}$=0.083 (the point $A$) is likely to be the
optimal one, since the closest maxima of $\max(p_1)$ are lower. As
predicted by the formula (10), if $\varepsilon_{field}$ approaches
0.14, the dashed line crosses the region where the transfer
probability substantially increases. At the point $B$ ($b$=3,
$\varepsilon_{field}$=0.14) the analytical value of $\max(p_1)$ =
0.991 correlates well with the numerical value of $\max(p_1)$ =
0.988. It should be noted that, despite of the appreciable
decrease in the transfer probability in comparison with the point
$A$, the value of transfer time $T_0$ = 250 for
$\varepsilon_{field}$ = 0.14 is about half of $T_0$ = 420 for
$\varepsilon_{field}$ = 0.083. We have also performed the
calculations of $\max(p_1)$ for the set of field and structure
parameters chosen from the high probability band containing the
points $C$ and $D$ (see Fig. 9). Again, the numerical data are
close to the analytical results, and $\max(p_1)$ grows
monotonically from 0.94 to 0.98, as we move from $C$ to $D$ along
the dashed broken line, whereas the transfer time decreases from
419 to 252. Going beyond the point $D$, we arrive at the crossing
region between this band and the vertical dashed line marking $b$
= 3, that covers the field energy interval around
$\varepsilon_{field}$ = 0.2 (not shown). For $\varepsilon_{field}$
= 0.2 we obtain $\max(p_1)$ =  0.986 that is smaller than the
value computed at the point $B$, but the corresponding transfer
time $T_0$ = 175  appears to be very short. It means that if we
are interested in fast electron transfer with moderate
reliability, it is recommended to take $\varepsilon_{field}$ = 0.2
instead of $\varepsilon_{field}$ = 0.083. At large values of the
parameters $\varepsilon_{field}$ and $b$, the probability of
successful electron transfer becomes quite low (dark area in a
right-up part of Fig. 9). This result is consistent with that
obtained earlier (see Refs. \cite{10} and \cite{11}) and is
explained in terms of the tunneling collapse.

At the final step, we have performed numerical simulations on the
electron dynamics in the nanostructure composed of $N$ QWs, where
the QW number varies from $N$ = 5 to $N$ = 20, and $b$ = 3. If $N$
is not large (5 $\le N \le$ 12), the dependencies of $\max(p_1)$
on the field energy $\varepsilon_{field}$ resemble those shown in
Fig. 7 and demonstrate quasiregular oscillations with the
amplitude increasing smoothly with $N$. For $N\ge$ 17, the plots
of $\max(p_1)$ versus $\varepsilon_{field}$ exhibit a "sawtooth"
profile with sharp deeps and peaks. Again, we make use of Eq. (10)
as the approximation for $\max(p_1)$, when the central state of
the excited subband serves as the transport state. We find that
the shapes of numerical and analytical curves of $\max(p_1)$
deviate substantially from each other, if $N$ approaches 20,
whereas the correspondence between the values of peak positions
remains surprisingly good. A general tendency is the shrinking of
the region, where the values of $\max(p_1)$ exceed 0.99. The last
pronounced peak of $\max(p_1)$ moves from $\varepsilon_{field}$ =
0.2 for $N$ = 6 to $\varepsilon_{field}$ = 0.08 for $N$ = 20. We
reveal, however, that the transfer times $T_0$ taken at the peaks
of $\max(p_1)$ are very close to each other for different $N$. In
Fig. 10 we present the dependencies of $T_0$ on
$\varepsilon_{field}$ for $N$ =15, 16, 17, and 18. The points of
interest, where 0.99 $\le\max(p_1)\le$ 0.999, are marked by open
(filled) squares for odd (even) $N$. It is easy to see that each
group of peaks, pertaining to different $N$, corresponds to a
narrow range of $T_0$ values. This observation enables us to make
an important conclusion that the electron transfer time in
multiwell nanostructure is not strictly defined by the number of
QW in the structure, as it might be expected, but displays rather
complex behavior conditioned by both the structure and field
parameters. As the final remark, we note that the optimal points
of the transfer probability lie in the middles of graph steps,
where the analytic expression $T_0=\pi/|\lambda_ {0,\,r}| \sqrt2$
yields very accurate values of the electron transfer time. Thus it
is expected that this simple formula can be used to estimate $T_0$
in the multiwell nanostructures regardless the QW number $N$.

\section{Conclusions}

In this paper, we have investigated theoretically the important
and very actual problem of coherent single-electron transfer
between two remote quantum wells in the semiconductor linear
nanostructure. Apart from wide interest to the fundamental
question addressing the possibility of reliable control over
individual charge carrier dynamics in the solid-state
low-dimensional structures, this study has a clear practical
application - i. e., the organization of indirect connections in
the ordered arrays of quantum bits.

For this purpose we make use of an auxiliary nanostructure
containing a single electron in the quantized part of its
conduction band. The resonant monochromatic laser pulse, applied
to the structure, drives the electron evolution in such a way
that, being initially localized in one quantum well, it comes onto
another well at the end of the pulse. This process is influenced
by many factors. In order to optimize the protocol - i. e., to
attain high probability of electron transfer in short time, one
should to choose carefully both structure and external field
parameters. As we have found, the maximum electron transfer
probability is a quasiperiodic function of the pulse strength and
the interlevel spacings in the excited subband. The choice of one
of excited states close to the subband center as the transport
state is substantiated by an observation that this state possesses
large ODT matrix element and is far enough from the others. It
provides high transfer characteristics such as speed and
selectivity.

Finally, it may be expected that the driving schemes relied upon
the stimulated Raman adiabatic passage (STIRAP) \cite{17,24,25}
together with optimal control techniques \cite{20} will enhance
the efficiency of optically-induced single-electron transfer in
the multiwell nanostructures.

\newpage
\centerline{\includegraphics[width=12cm]{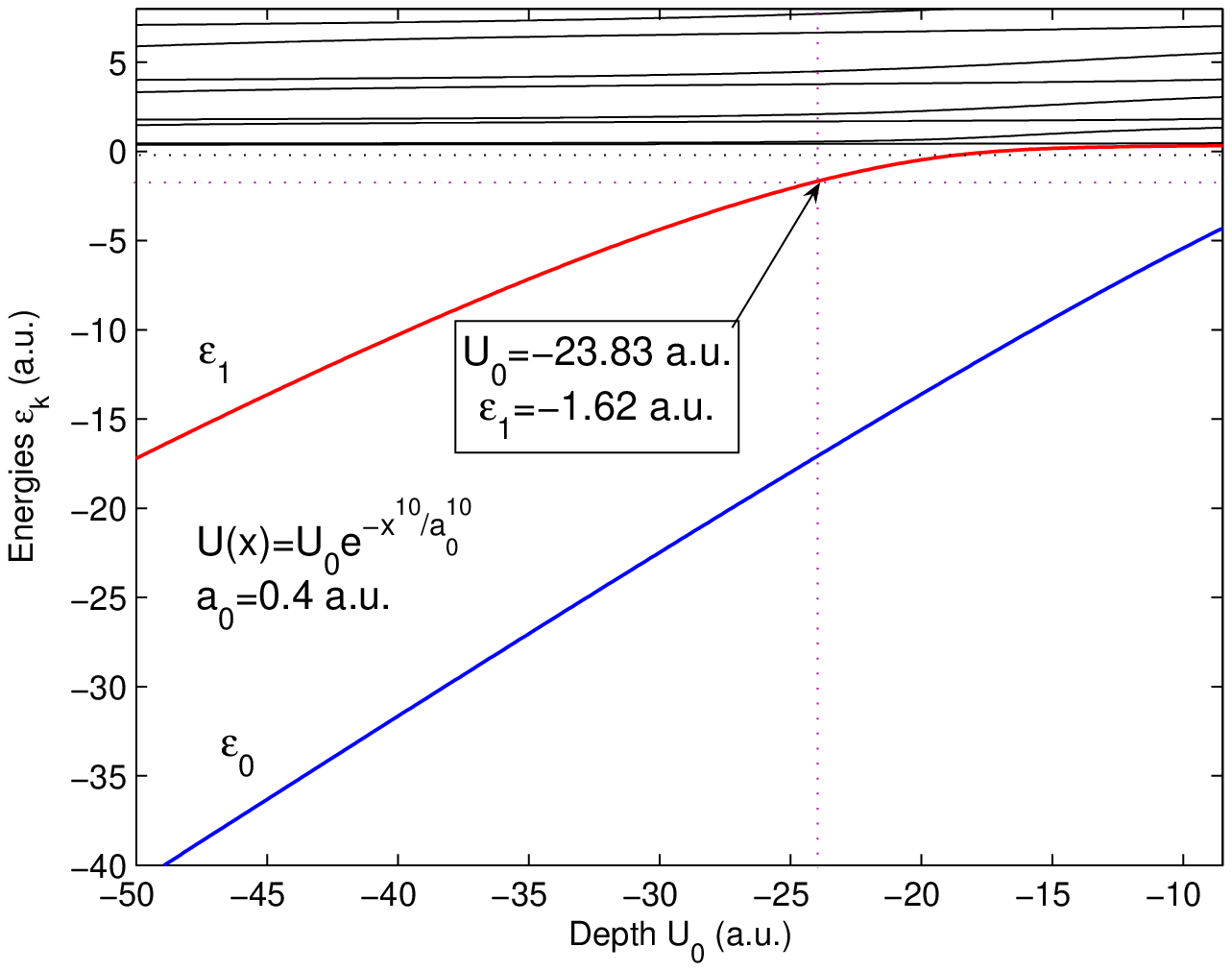}}
 \centerline{Fig. 1 (a)}

\centerline{\includegraphics[width=12cm]{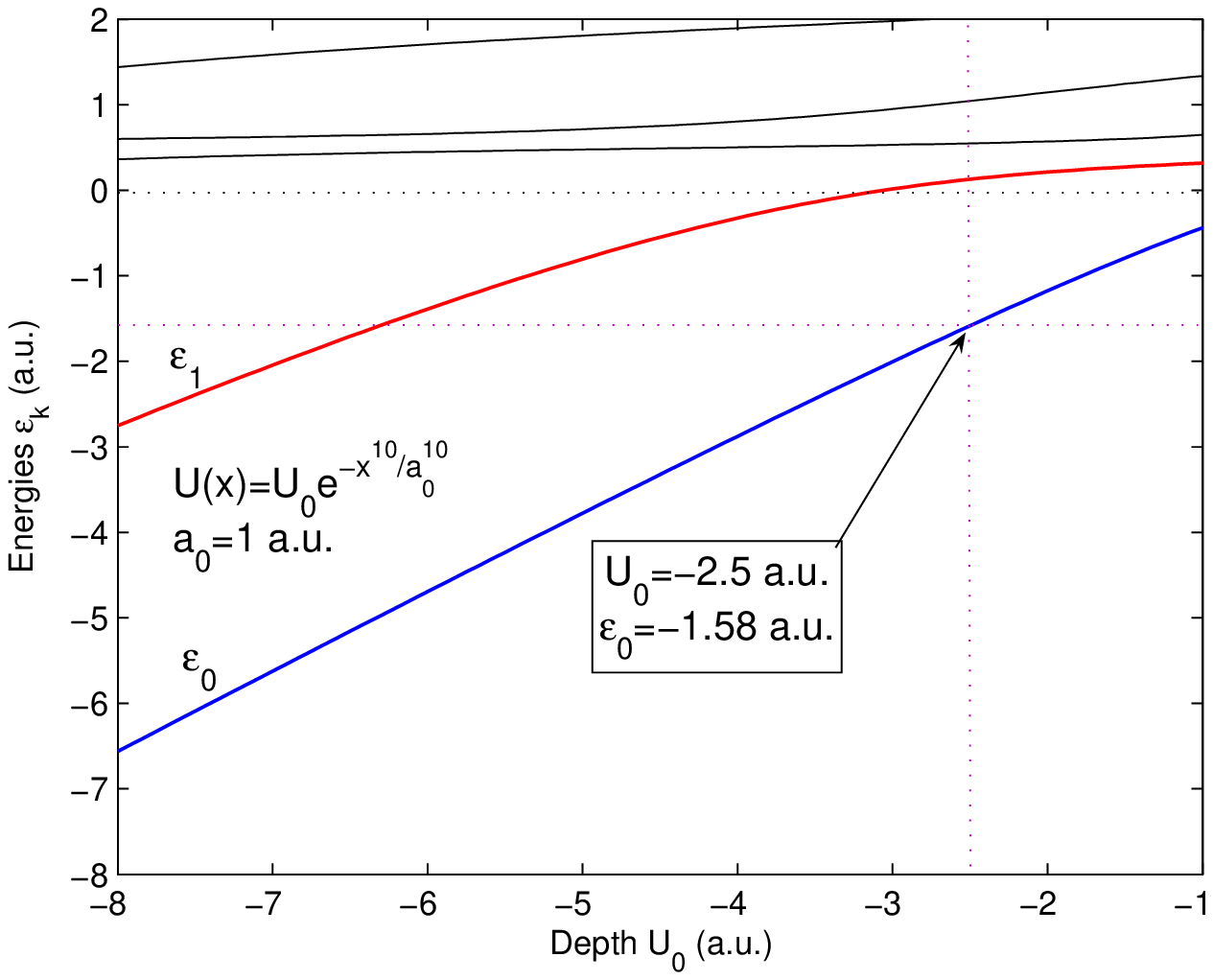}}
 \centerline{Fig. 1 (b)}

\newpage
\centerline{\includegraphics[width=12cm]{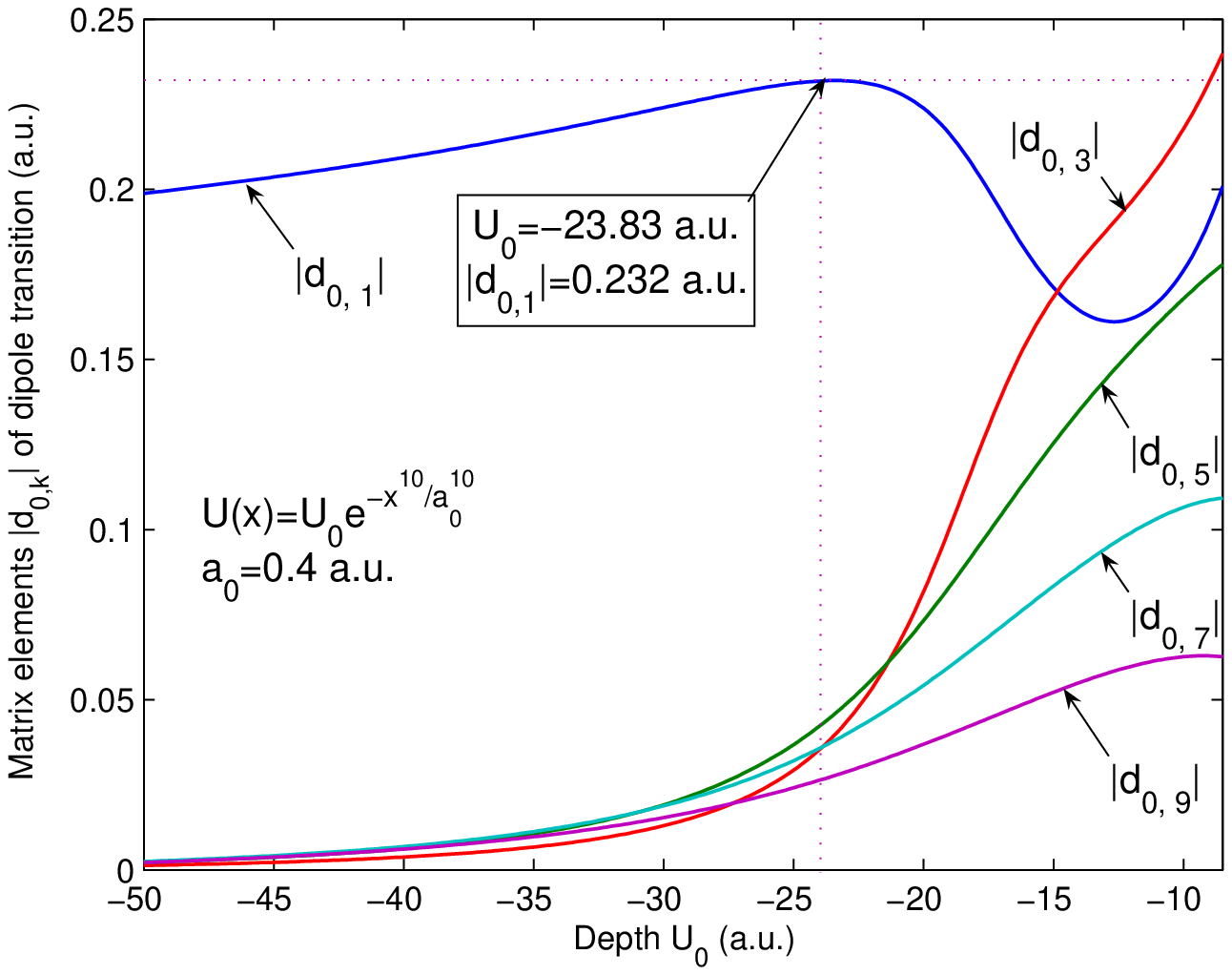}}
 \centerline{Fig. 2 (a)}

\centerline{\includegraphics[width=12cm]{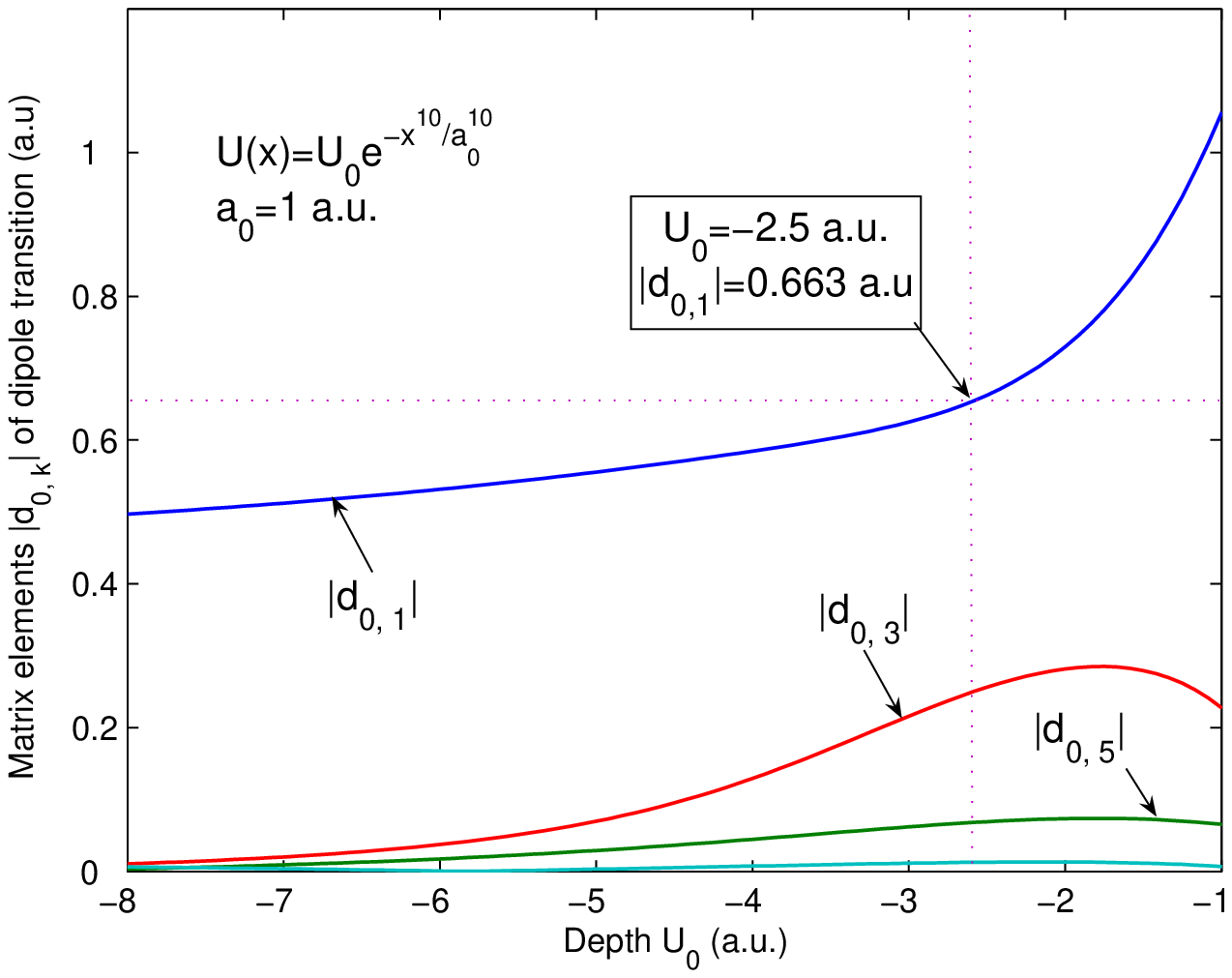}}
 \centerline{Fig. 2 (b)}

 \newpage
\centerline{\includegraphics[width=12cm]{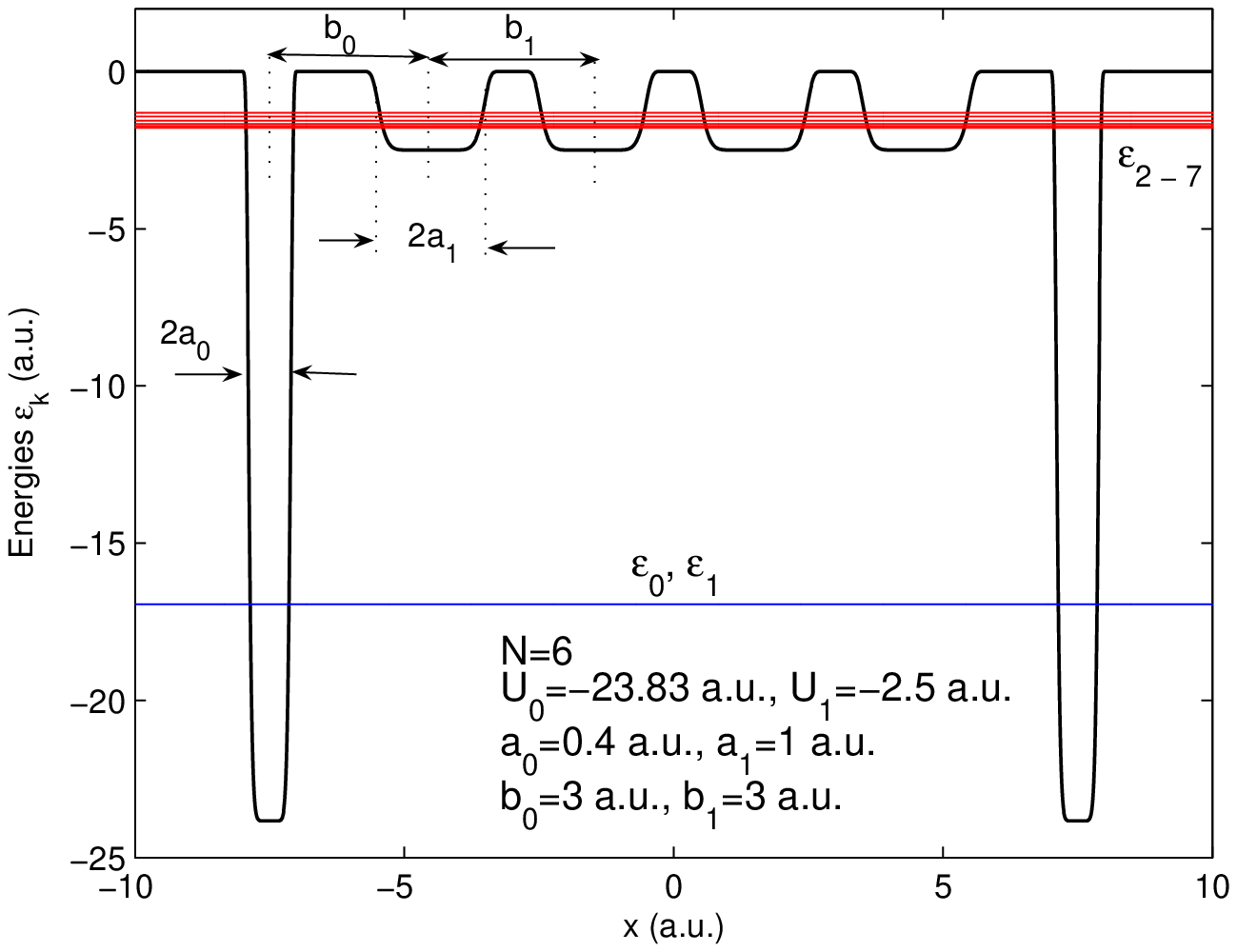}}
 \centerline{Fig. 3 (a)}

\centerline{\includegraphics[width=12cm]{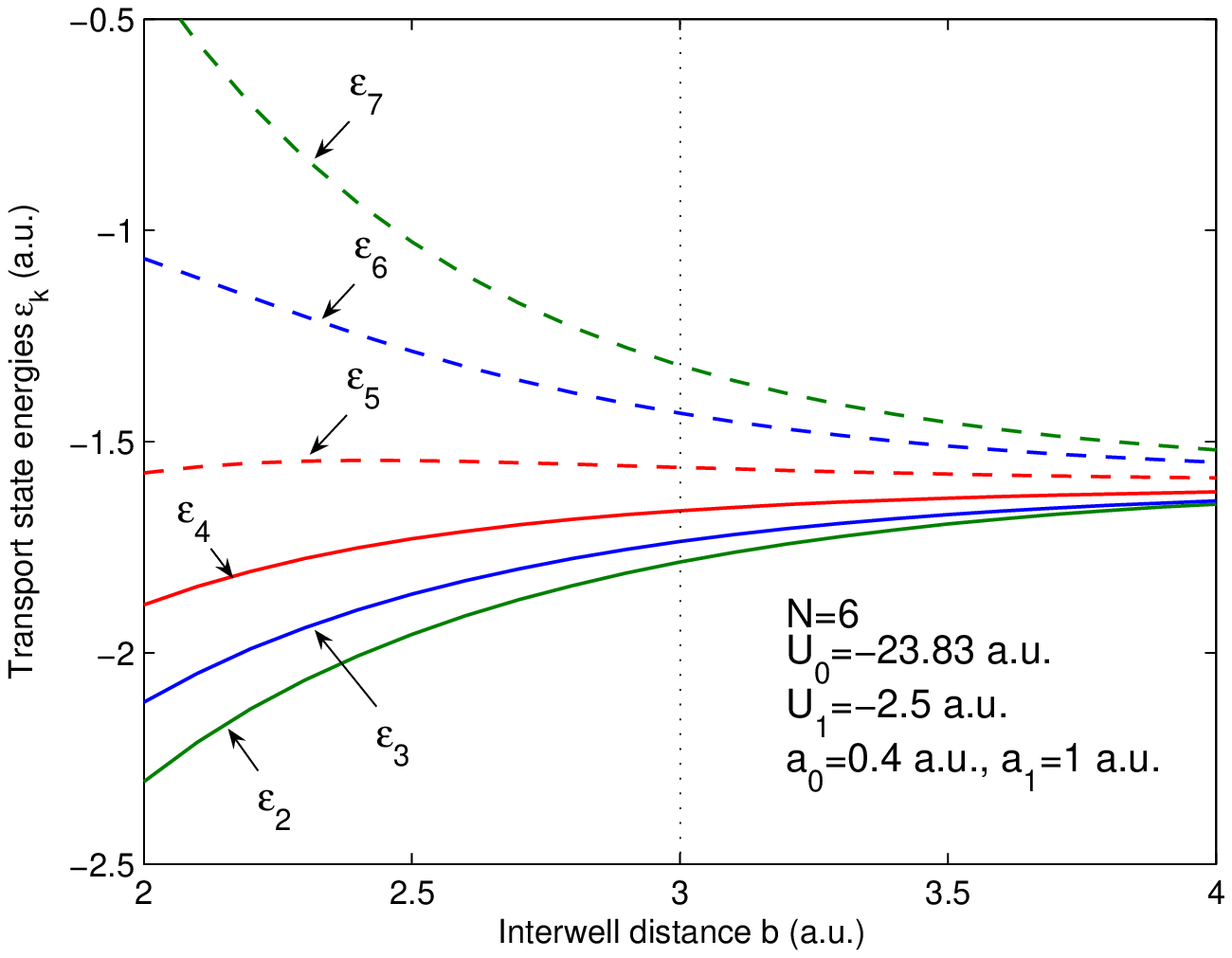}}
 \centerline{Fig. 3 (b)}

\newpage
\centerline{\includegraphics[width=12cm]{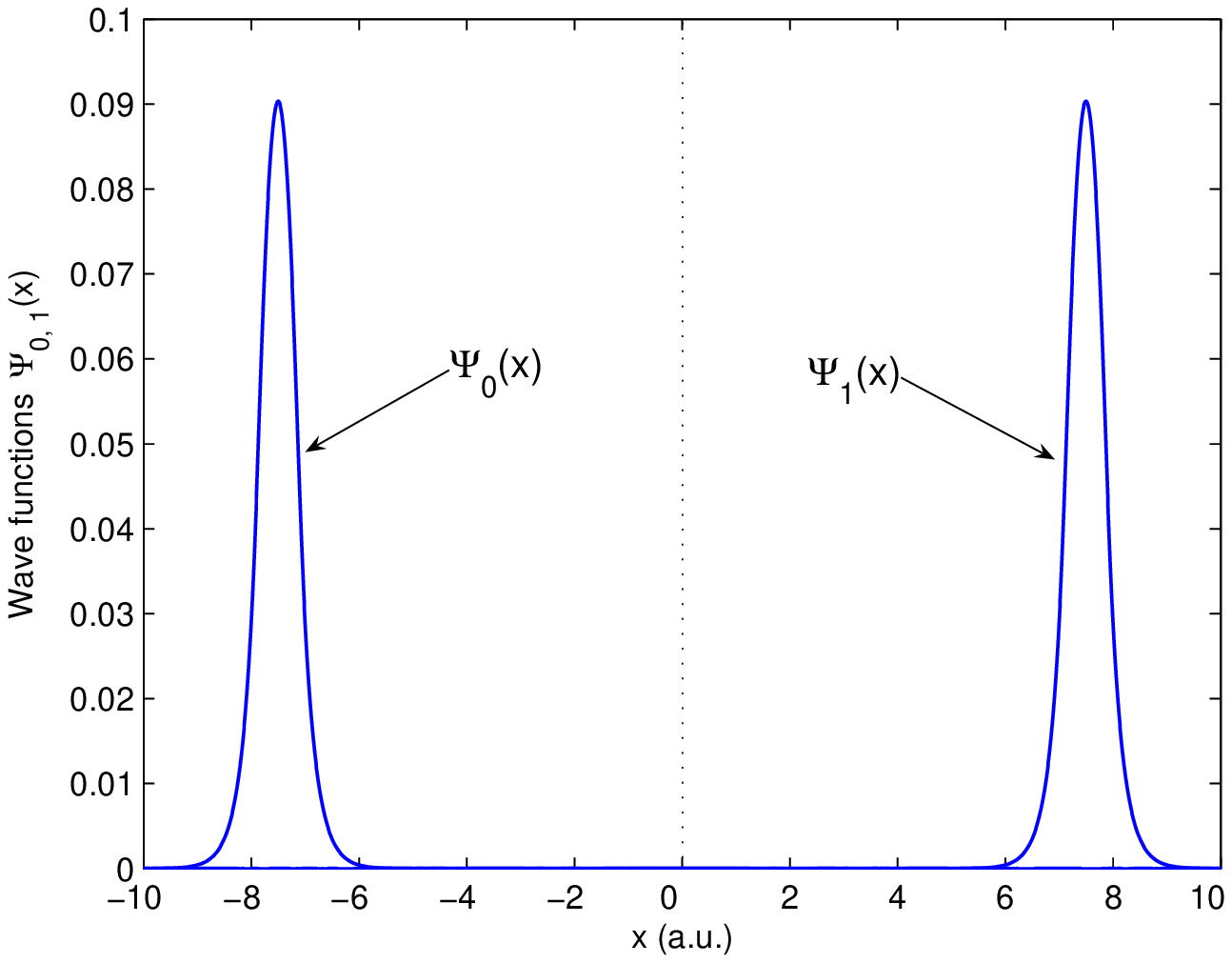}}
 \centerline{Fig. 4 (a)}

\centerline{\includegraphics[width=12cm]{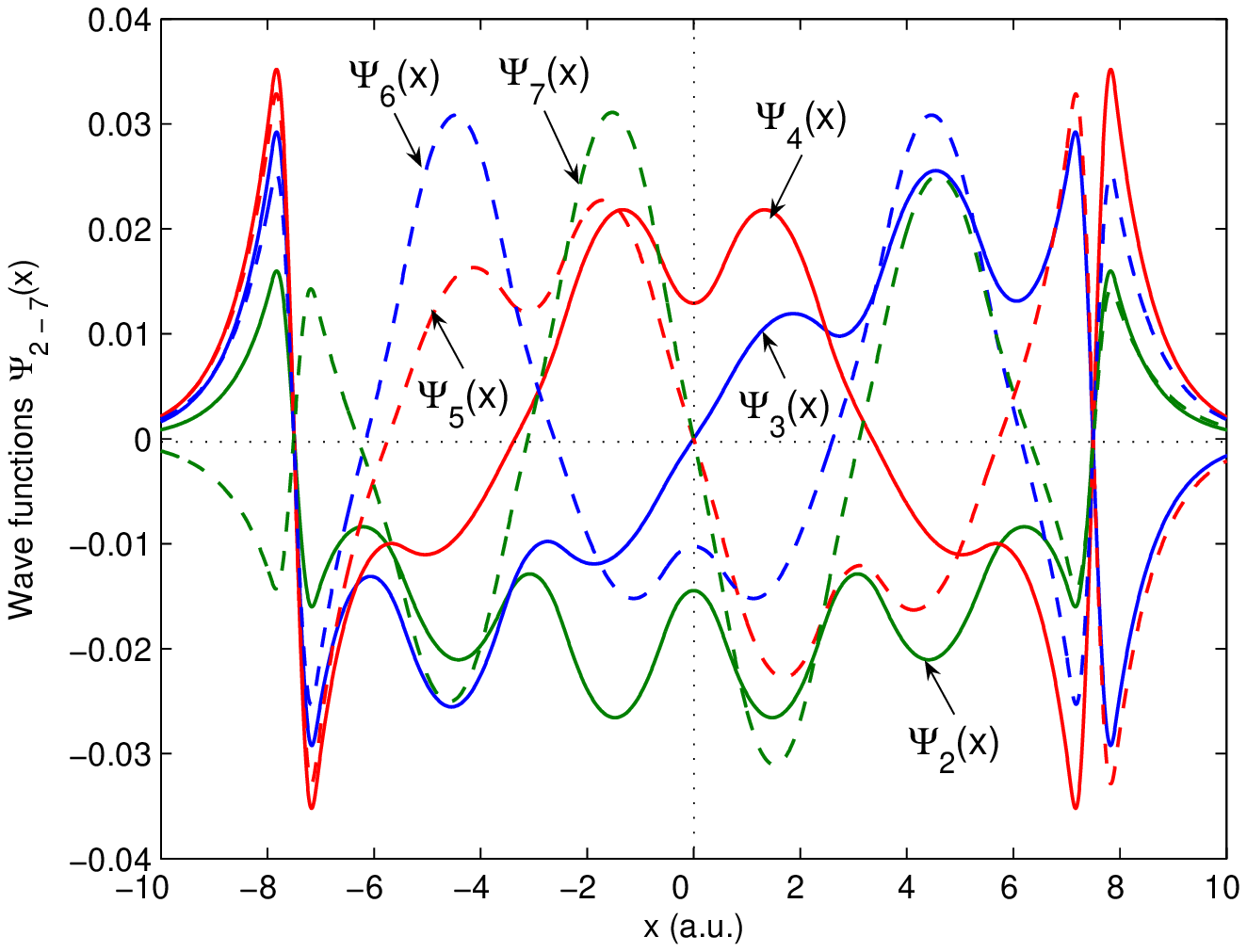}}
 \centerline{Fig. 4 (b)}

\newpage
\centerline{\includegraphics[width=12cm]{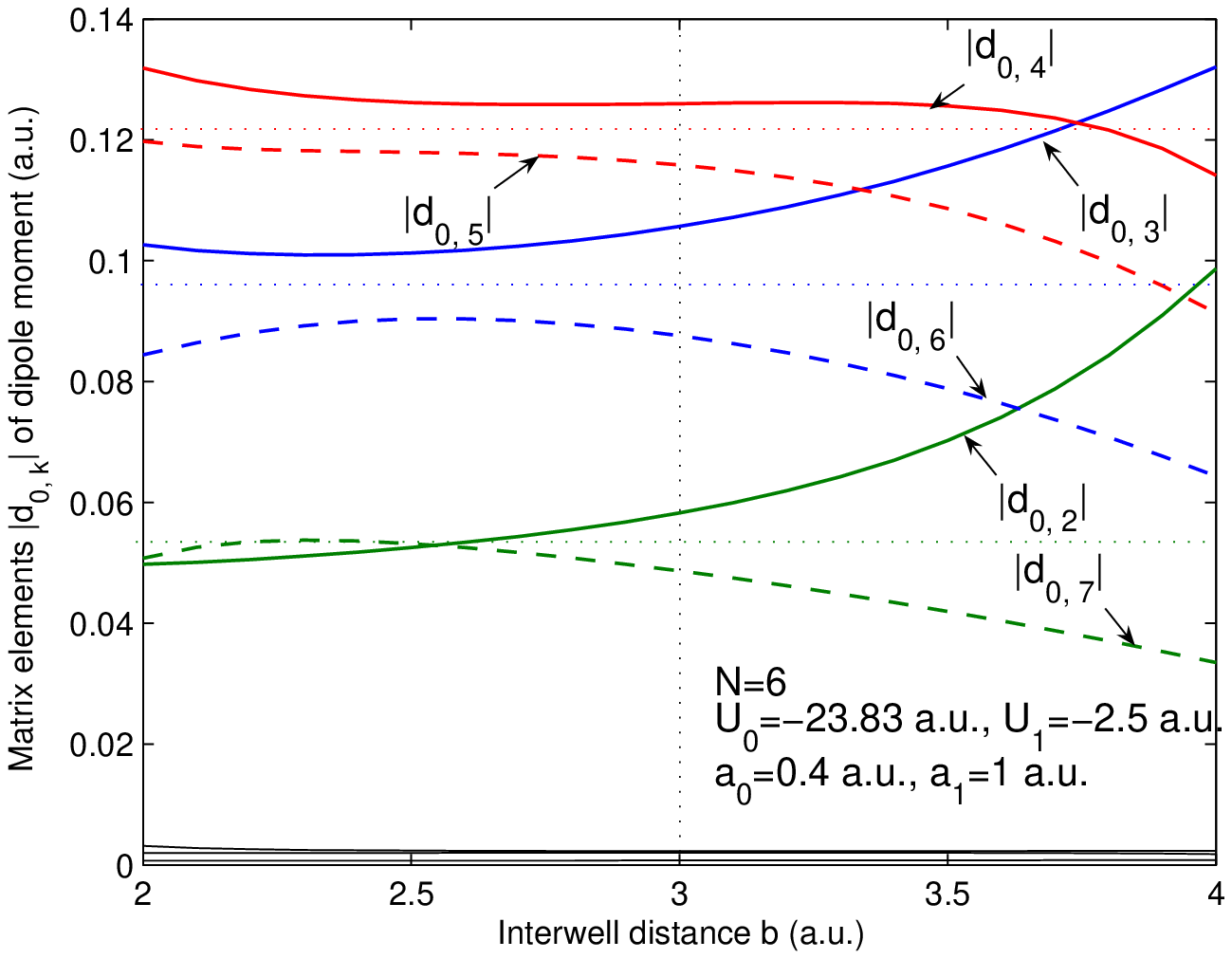}}
 \centerline{Fig. 5 (a)}

\centerline{\includegraphics[width=12cm]{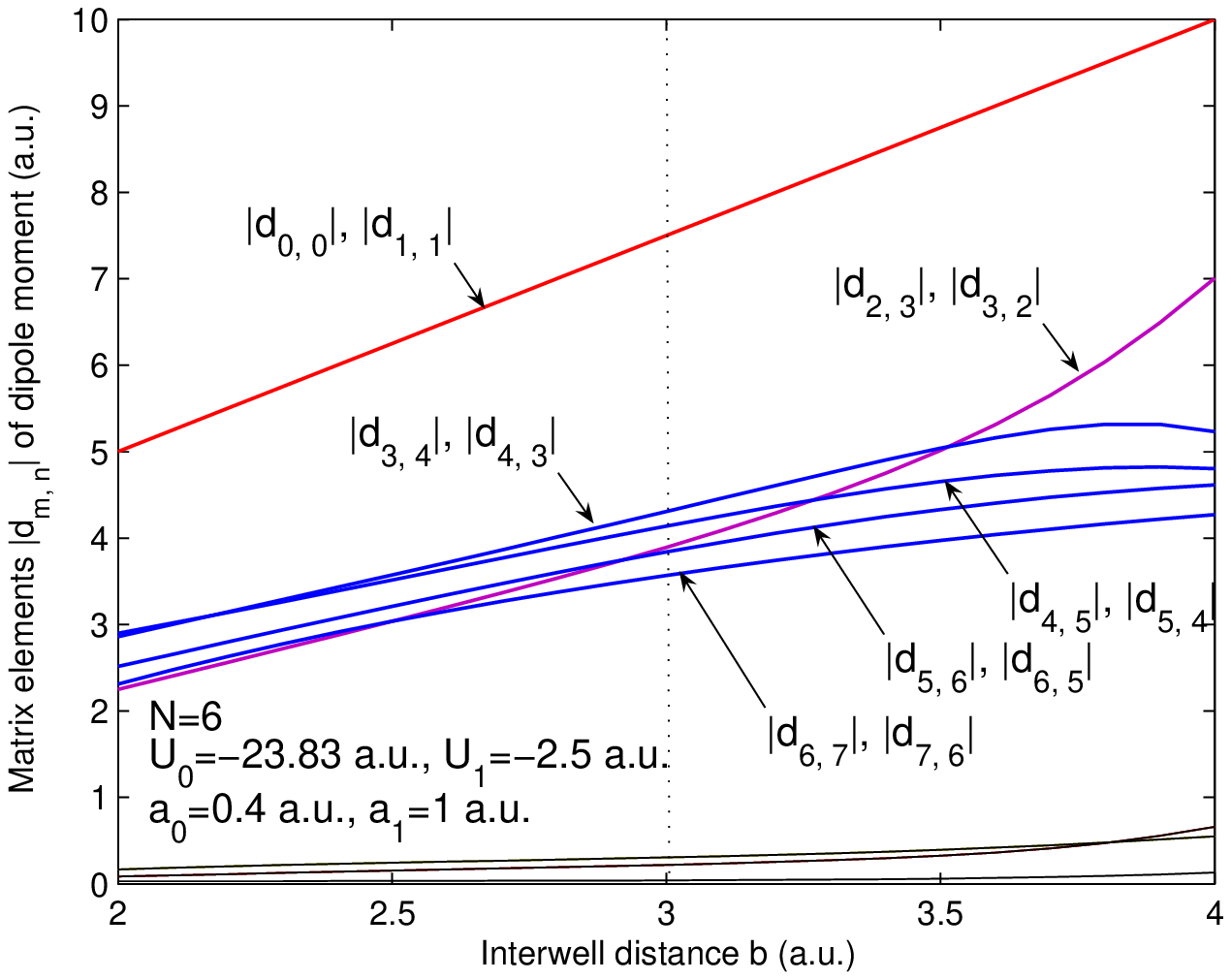}}
 \centerline{Fig. 5 (b)}

\newpage
\centerline{\includegraphics[width=12cm]{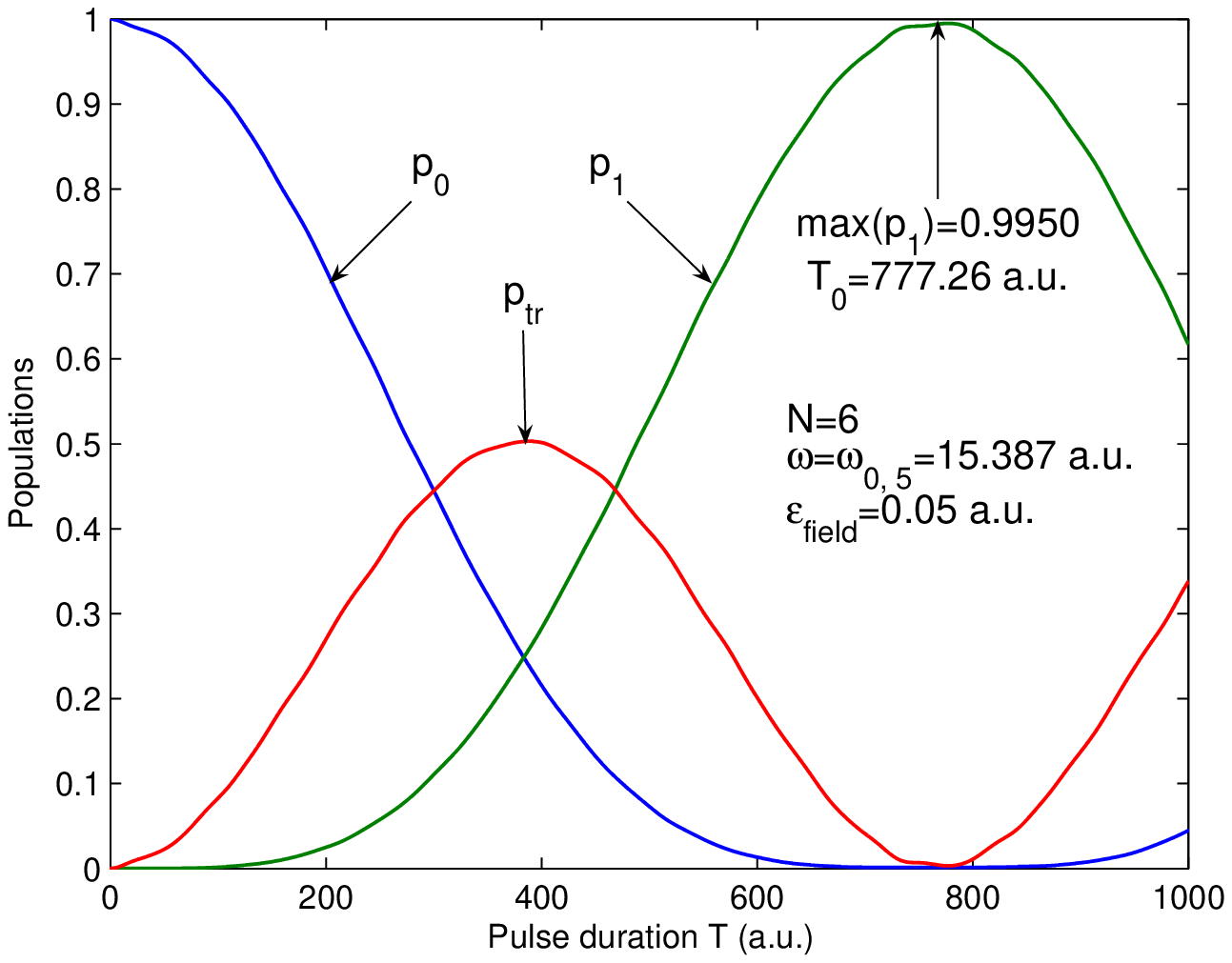}}
 \centerline{Fig. 6}

\centerline{\includegraphics[width=12cm]{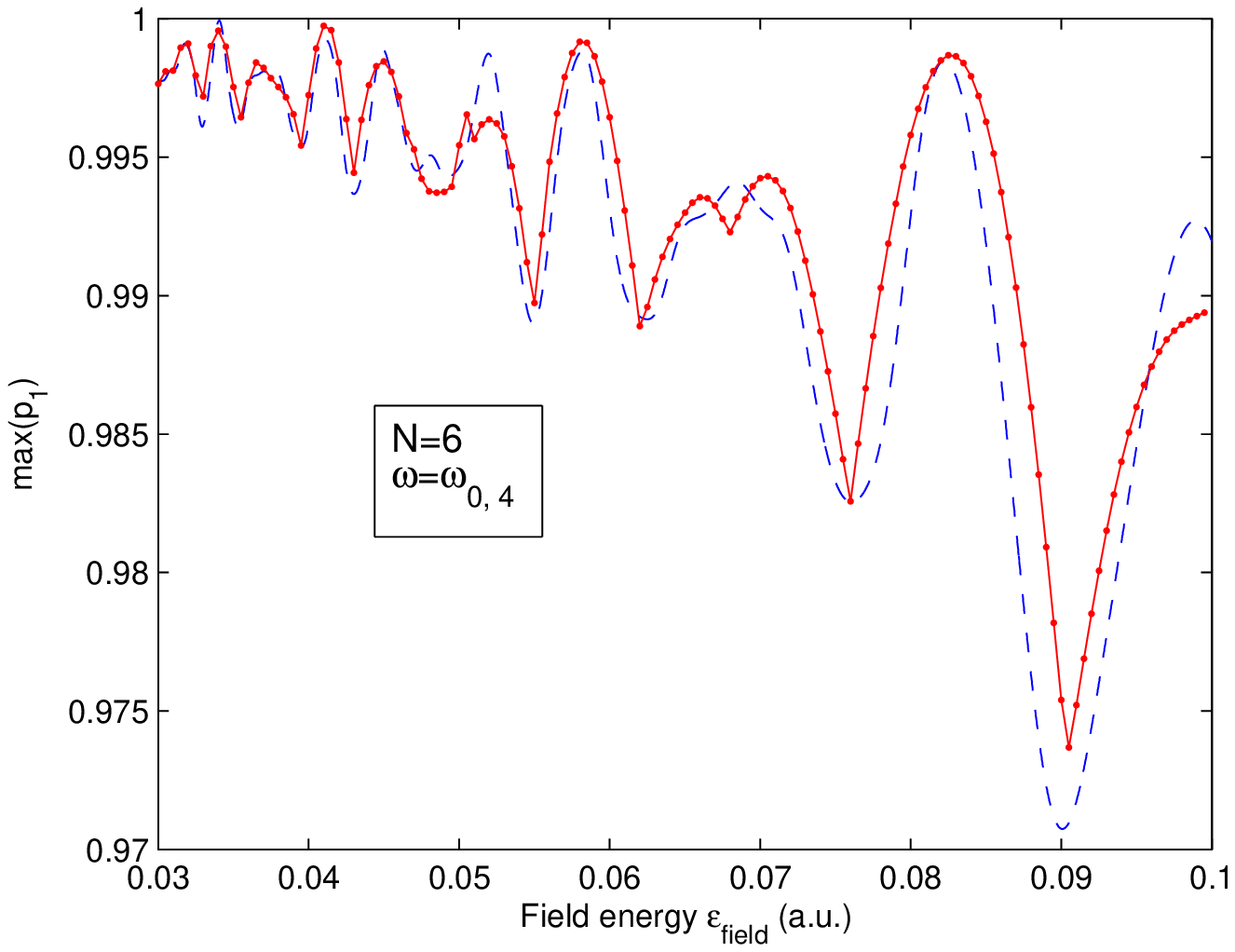}}
 \centerline{Fig. 7 (a)}

\newpage
\centerline{\includegraphics[width=12cm]{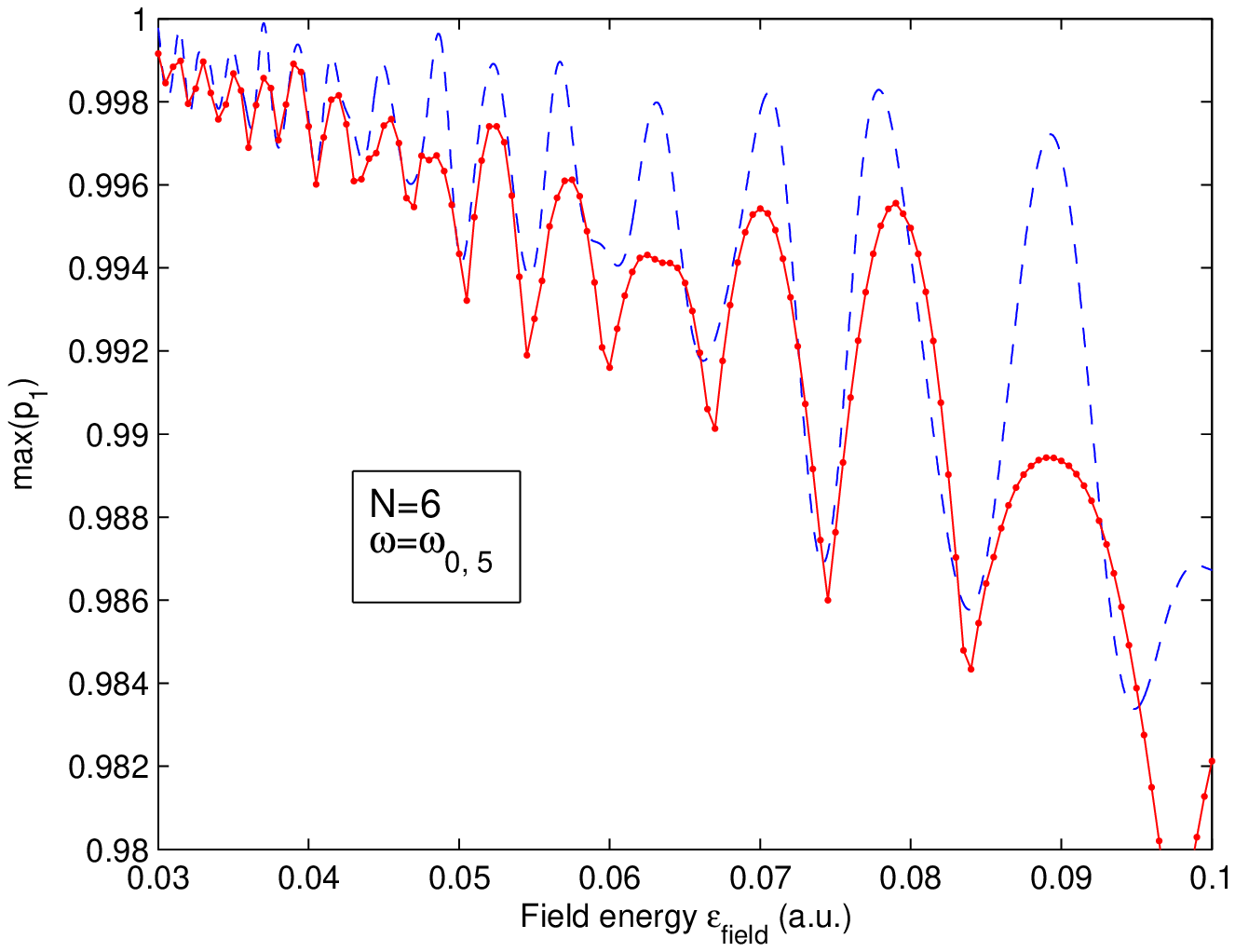}}
 \centerline{Fig. 7 (b)}

\centerline{\includegraphics[width=12cm]{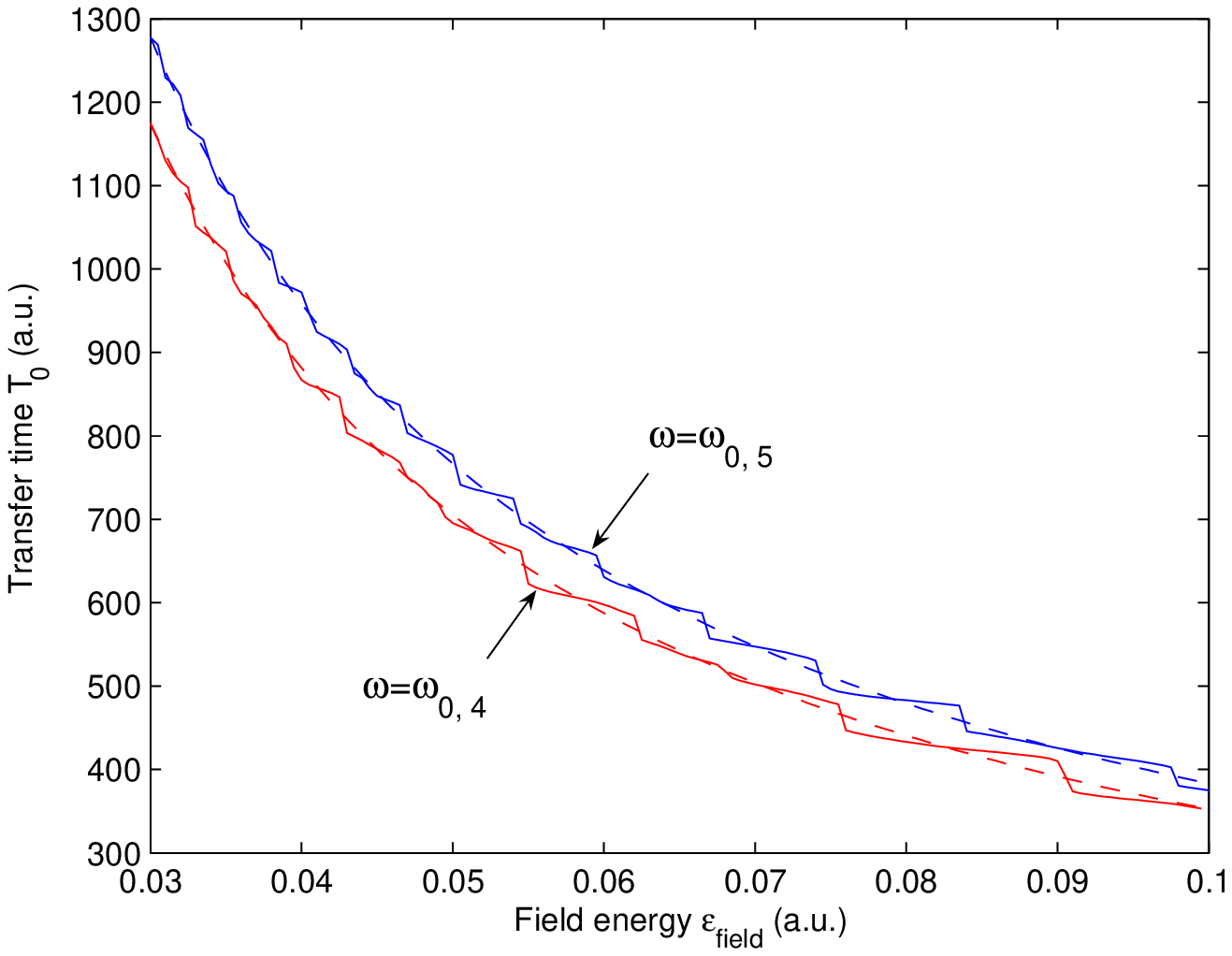}}
 \centerline{Fig. 8}

\newpage
\centerline{\includegraphics[width=12cm]{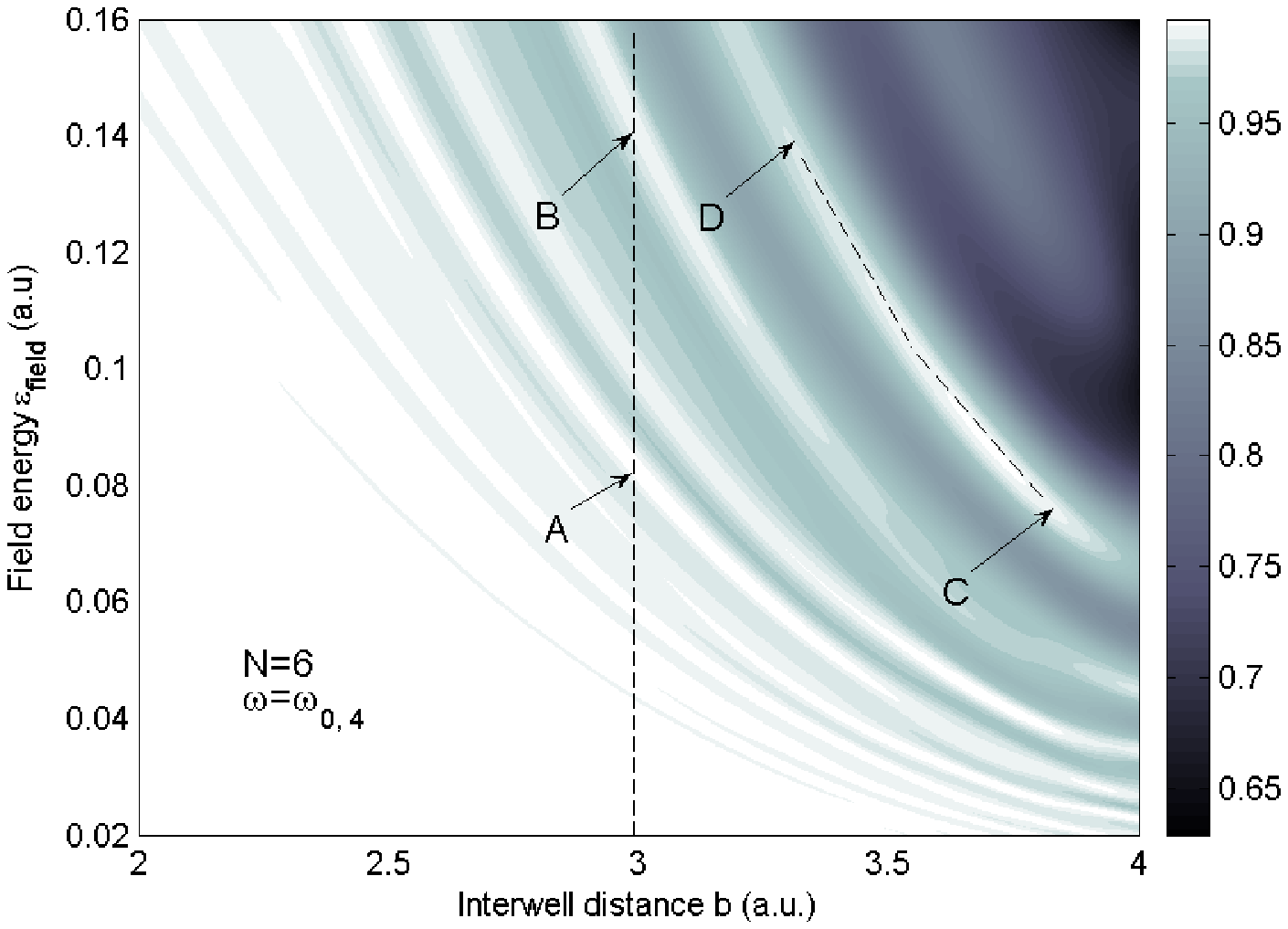}}
 \centerline{Fig. 9}

\centerline{\includegraphics[width=12cm]{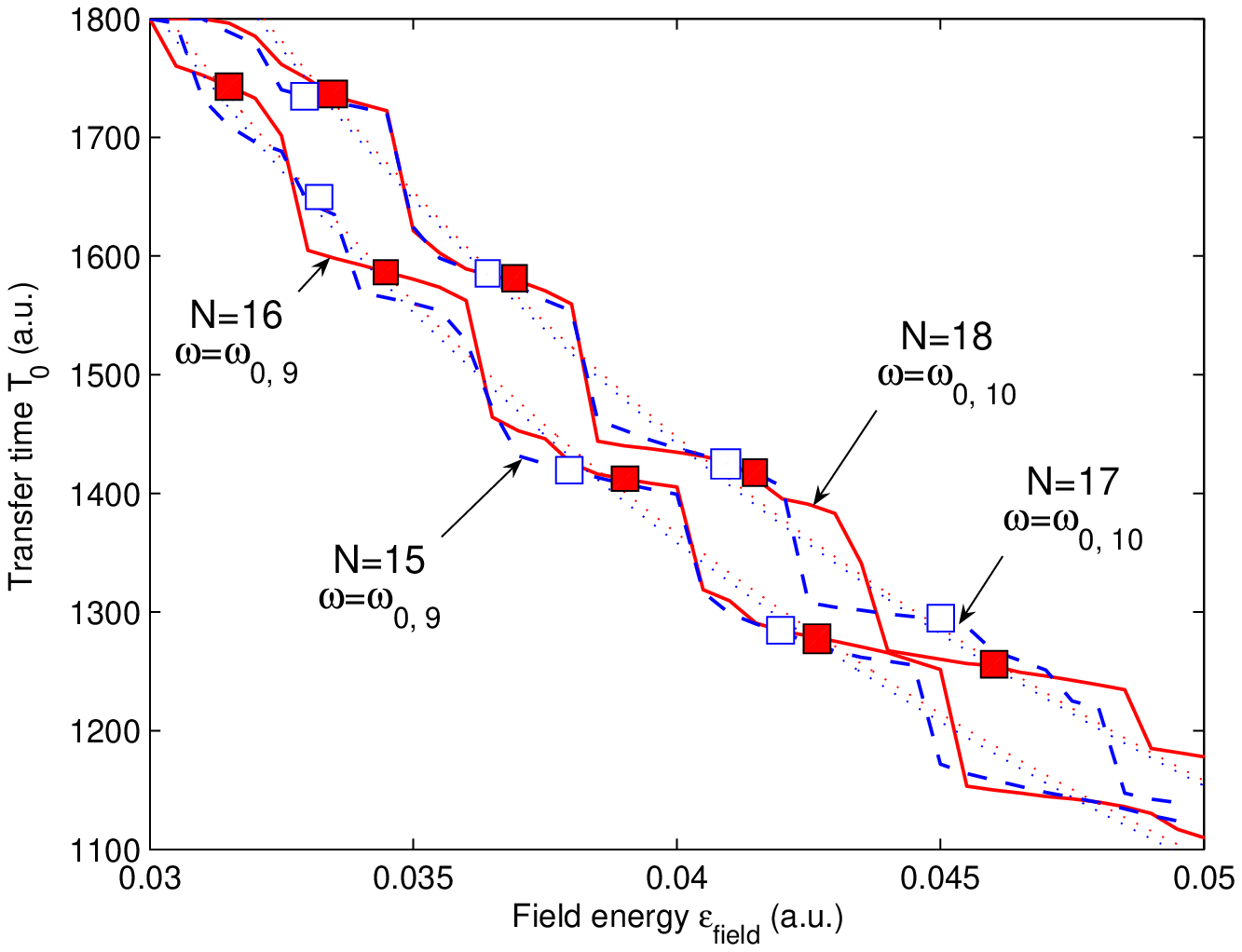}}
 \centerline{Fig. 10}

\newpage

\Figures

\Figure{(color online). The dependencies of electronic
eigenenergies $\varepsilon_k$ of low-lying eigenstates of a single
PE quantum well on the potential depth $U_0$ for a) the edge
quantum well and  b) the internal quantum well. The well
parameters used in further numerical calculations are shown in the
insets.}

\Figure{(color online). The values $|d_{0, k}|$ of matrix elements
of optical dipole transitions, connecting the ground state $\left|
0\right\rangle$ and the low-lying excited states $\left|
k\right\rangle$ for a) the edge quantum well and  b) the internal
quantum well.}

\Figure{(color online). The six-well one-dimensional nanostructure
placed in the central part of the large quantum well ($L$ = 100
a.u.). a) Potential profile $U(x)$ of the structure in the case of
$b_0=b_1$ = 3 a.u. and the energy levels $\varepsilon_k$ from
ground ($k$=0, 1) and first-excited ($k$ = 2 - 7) subbands. b) The
dependencies of the excited state energies $\varepsilon_k$ ($k$ =
2 - 7) on the interwell distance $b=b_0=b_1$. Vertical dotted line
marks the value $b$ = 3 a.u. at which the most of numerical
simulations are performed. Other parameters are given in the
plot.}

\Figure{(color online). The plots of the electron wave functions
for low-lying states of the six-well nanostructure (in arbitrary
units) vs the coordinate $x$ at $b$=3 a.u. a) Ground-state wave
functions $\Psi_0 \left( x \right)$ and $\Psi_1 \left( x \right)$;
b) Excited-state wave functions $\Psi_k \left( x \right)$, $k$ = 2
- 7. }

\Figure{(color online). The absolute values of ODT matrix elements
of the six-well nanostructure as the functions of interwell
distance $b$. a) Intersubband matrix elements $|d_{0, k}|$ . The
horizontal dotted lines designate the values of $|d_{0, k}|$
calculated in the TB approximation. b) Intrasubband  matrix
elements $|d_{m, n}|$.}

\Figure{(color online). The populations $p_0$ and $p_1$ of the
ground states $\left| 0 \right\rangle $ and $\left| 1
\right\rangle $, and the total population $p_{tr}$ of excited
states states $\left| k \right\rangle $, $k\ge$2, vs the pulse
duration $T$. The structure and pulse parameters as well as the
value of the transfer probability $\max(p_1)$ are shown on the
plot.}

\Figure{(color online). The maximum transfer probability
$\max(p_1)$ of electron in the six-well nanostructure plotted as a
function of the field energy $\varepsilon_{field}$ for two choices
of the transport state: a) $r$=4 and b) $r$=5. The numerical
solutions are presented by the solid curves (the calculated points
are denoted by circles); the approximations of Eq. (10) are shown
by the dashed curves.}

\Figure{(color online). The transfer time $T_0$ between the ground
states of the six-well structure vs the field energy
$\varepsilon_{field}$ for two choices of the transport state,
$r$=4 and $r$=5. The plots visualizing the approximation
$T_0=\pi/|\lambda_ {0,\,r}| \sqrt2$ are presented by dashed
curves.}

\Figure{(color online). The transfer probability $\max(p_1)$, Eq.
(10), as a function of both the interwell distance $b$ and the
field energy $\varepsilon_{field}$. The dashed lines denote the
paths along which the numerical simulations were performed.}

\Figure{(color online). The transfer times $T_0$ vs the field
energy $\varepsilon_{field}$ for the QW numbers $N$ =15, 17 (blue
dashed curves, open squares) and $N$ =16, 18 (red solid curves,
filled squares). The analytic data are shown by thin dotted
curves.}

\end{document}